\documentclass[aip, pop, amsfonts, floatfix, amssymb, twocolumn, amsmath, reprint,nofootinbib]{revtex4-1}

\usepackage{times}
\usepackage{bm}
\usepackage{graphicx}
\usepackage{graphics}
\usepackage{epsfig}
\usepackage{epstopdf}
\usepackage[colorinlistoftodos]{todonotes}
\usepackage{xcolor}

\usepackage{booktabs}
\usepackage{amsmath}
\usepackage{tabularx}
\usepackage{xcolor}
\usepackage[colorlinks=true, allcolors=blue, breaklinks]{hyperref}
\usepackage{mwe}

\newcommand*{\citen}[1]{%
  \begingroup
    \romannumeral-`\x 
    \setcitestyle{numbers}%
    \cite{#1}%
  \endgroup
}

\begin{document}

\title{Gyrokinetic GENE simulations of DIII-D near-edge L-mode plasmas}

\author{T. F.~Neiser}
\email{tomneiser@physics.ucla.edu}
\affiliation{Department of Physics and Astronomy, University of California, Los Angeles, CA 90095$\text{,}$ USA}

\author{F.~Jenko}
\affiliation{Department of Physics and Astronomy, University of California, Los Angeles, CA 90095$\text{,}$ USA}
\affiliation{Max Planck Institute for Plasma Physics, Boltzmannstr. 2, 85748 Garching, Germany}
\affiliation{Institute for Fusion Studies, University of Texas at Austin, Austin, TX 78712, USA}

\author{T. A.~Carter}
\affiliation{Department of Physics and Astronomy, University of California, Los Angeles, CA 90095$\text{,}$ USA}

\author{L.~Schmitz}
\affiliation{Department of Physics and Astronomy, University of California, Los Angeles, CA 90095$\text{,}$ USA}

\author{D.~Told}
\affiliation{Max Planck Institute for Plasma Physics, Boltzmannstr. 2, 85748 Garching, Germany}

\author{G.~Merlo}
\affiliation{Institute for Fusion Studies, University of Texas at Austin, Austin, TX 78712, USA}

\author{A.~Ba\~n\'on Navarro}
\affiliation{Max Planck Institute for Plasma Physics, Boltzmannstr. 2, 85748 Garching, Germany}

\author{P.~C.~Crandall}
\affiliation{Department of Physics and Astronomy, University of California, Los Angeles, CA 90095$\text{,}$ USA}

\author{G.~R.~McKee}
\affiliation{Department of Physics, University of Wisconsin-Madison, Madison, WI 53706, USA}

\author{Z.~Yan}
\affiliation{Department of Physics, University of Wisconsin-Madison, Madison, WI 53706, USA}


\begin{abstract}
We present gyrokinetic simulations with the GENE code addressing the near-edge region of an L-mode plasma in the DIII-D tokamak. At radial position $\rho=0.80$, simulations with the ion temperature gradient increased by $40\%$ above the nominal value give electron and ion heat fluxes that are in simultaneous agreement with the experiment. This gradient increase is consistent with the combined statistical and systematic uncertainty $\sigma$ of the Charge Exchange Recombination Spectroscopy (CER) measurements at the $1.6 \sigma$ level. Multi-scale simulations are carried out with realistic mass ratio and geometry for the first time in the near-edge. These multi-scale simulations suggest that the highly unstable ion temperature gradient (ITG) modes of the flux-matched ion-scale simulations suppress electron-scale transport, such that ion-scale simulations are sufficient at this location. At radial position $\rho=0.90$, nonlinear simulations show a hybrid state of ITG and trapped electron modes~(TEMs), which was not expected from linear simulations. The nonlinear simulations reproduce the total experimental heat flux with the inclusion of $\mathbf{E} \times \mathbf{B}$ shear effects and an increase in the electron temperature gradient by $\sim 23\%$. This gradient increase is compatible with the combined statistical and systematic uncertainty of the Thomson scattering data at the $1.3 \sigma$ level. These results are consistent with previous findings that gyrokinetic simulations are able to reproduce the experimental heat fluxes by varying input parameters close to their experimental uncertainties, pushing the validation frontier closer to the edge region.
\end{abstract}

\maketitle

\section{Introduction}\label{intro}
In order to improve the energy confinement time of magnetic fusion experiments, a thorough understanding of turbulent transport is necessary. The main carriers of cross-field transport in tokamaks are gyroradius-scale microinstabilities with scale lengths much smaller than the machine size. These microinstabilities are physically driven by electron and ion temperature gradients and density gradients. Since microturbulence is suspected to reduce energy confinement time with an increase in heating power, its mitigation is of intrinsic interest to the fusion community. 

Paradoxically, a state of improved confinement arises together with steepened gradients in the edge region when the heating power is increased above a certain threshold power,~$P_{\rm th}$. This transition from low confinement mode (L-mode) to high confinement mode (H-mode) was first discovered at the ASDEX tokamak in 1982 and has since been reproduced in all major tokamaks \cite{Wagner82, Wagner}. H-mode is the favored operational regime for nuclear fusion reactors and ITER. However, finding a self-consistent description of the L-H transition is a major unsolved problem \cite{Wagner, ConnorWilson, Batchelor07}. An important first step towards understanding the L-H transition is to correctly describe L-mode plasmas in the near-edge and edge regions. This is also important for ITER, which will be in L-mode operation during plasma current ramp-up and ramp-down phases; correctly predicting the L-mode profiles is important for vertical stabilization of the plasma during these phases \cite{Casper10, Parail13}. This motivates a study of microinstabilities in the L-mode near-edge just before an L-H transition. 

Gyrokinetic theory provides an accurate description of microturbulence in magnetically confined plasmas. Here, the assumptions of high background magnetic field, low frequencies relative to the ion cyclotron frequency and small fluctuation amplitudes typically apply \cite{Krommes12}. Gyrokinetic codes such as GKV\cite{GKV1,GKV2}, GEM \cite{Chen03,Chen07}, GYRO\cite{CandyWaltz} and GENE\cite{GENE} have been in good agreement with the experiment and with each other in the core of both L-mode\cite{Holland09,RhodesNF11, Goerler14} and H-mode\cite{Holland11, Holland12, Banon15} plasmas. Similarly, the near-edge region of H-mode plasmas has been successfully modeled \cite{Holland11, Holland12, Hatch15}. However, these codes have been in occasional disagreement in the near-edge of L-mode plasmas. For example, simulations have shown an underprediction of heat transport of $\sim 7$ for GYRO \cite{Holland09} and $\sim 2$ for GENE \cite{Goerler14} for nominal parameters of DIII-D discharge \#128913. This has raised fears of an apparent systematic shortfall of heat flux predictions \cite{Holland09, RhodesNF11}. However, these fears have recently been reduced \cite{DTold13, Goerler14, Waltz17, HowardNF13, HowardNF16}. For example, increasing the ion temperature gradient within the experimental error bars has produced flux-matched simulations with GENE \cite{Goerler14}. Recent GYRO simulations of a different L-mode discharge, namely DIII-D \#101391, have revisited this shortfall problem and found good agreement with experiment \cite{Waltz17}. Similarly, the CGYRO code \cite{Candy16} matches the experimental heat flux of this discharge \cite{Waltz17}, as does the GENE code. Studies with GYRO on the Alcator C-Mod tokamak have also been in good agreement with experiment near the edge region of L-modes, with the use of multi-scale simulations in some cases \cite{HowardNF13, HowardNF16}. These gyrokinetic validation exercises reduce fears that the shortfall is a universal feature of near-edge L-mode plasmas. However, these fears have not been completely removed yet. Moreover, the role of microturbulence and its interactions on multiple scales is still poorly understood in the near-edge. Therefore, there exists a continued need for code validation and microturbulence characterization in the near-edge of L-mode plasmas. 

To address these issues, we present a gyrokinetic validation study of a DIII-D near-edge L-mode plasma just before an L-H transition with the gyrokinetic turbulence code GENE. Our primary finding is that gyrokinetic simulations are able to match the heat-flux in the near-edge\footnote[2]{We loosely define the near-edge as $0.80\leq \rho \leq 0.96$, where $\rho=~(\Phi/\Phi_{\rm edge})^{1/2}$ is the toroidal flux radius and $\Phi_{\rm edge}$ is the toroidal flux at the separatrix.} of the L-mode plasma at $\rho=0.8$ and $\rho=0.9$ within the uncertainty of the experiment at the $1.6 \sigma$ and $1.3 \sigma$ levels, respectively. In the course of this validation study, we make three secondary findings. First, current heuristic rules for the relevance of multi-scale effects appear to be on the cautious side; multi-scale simulations at $\rho=0.80$ suggest that single-scale simulations can be sufficient in a scenario when multi-scale effects are expected, which could increase the realm of applicability of single-scale simulations. 
Second, the effect of edge $\bm{E}\times \bm{B}$ shear is found to be already important in the near-edge at $\rho=0.90$, which was unexpected. Third, nonlinear simulations at $\rho=0.90$ find a hybrid ion temperature gradient (ITG)/ trappend electron mode (TEM) scenario, which was not obvious from linear simulations. This successful validation exercise helps push the gyrokinetic validation frontier closer to the L-mode edge region. 

This paper is structured as follows. We discuss the experimental data in section~\ref{exp} and the gyrokinetic simulation method in section~\ref{meth}. In section~\ref{linsim08}, we present linear and nonlinear simulation results at radial position $\rho=0.80$. In section~\ref{edge}, we present linear and nonlinear simulation results closer to the edge at $\rho=0.90$. We reflect on the limitations and implications of our results in section~\ref{disc}, summarize the present work in section~\ref{sum} and outline future work in section~\ref{future}. 

\section{Summary of Experimental Data \label{exp}}

\begin{figure}[!htb]\includegraphics[width=0.457\textwidth]{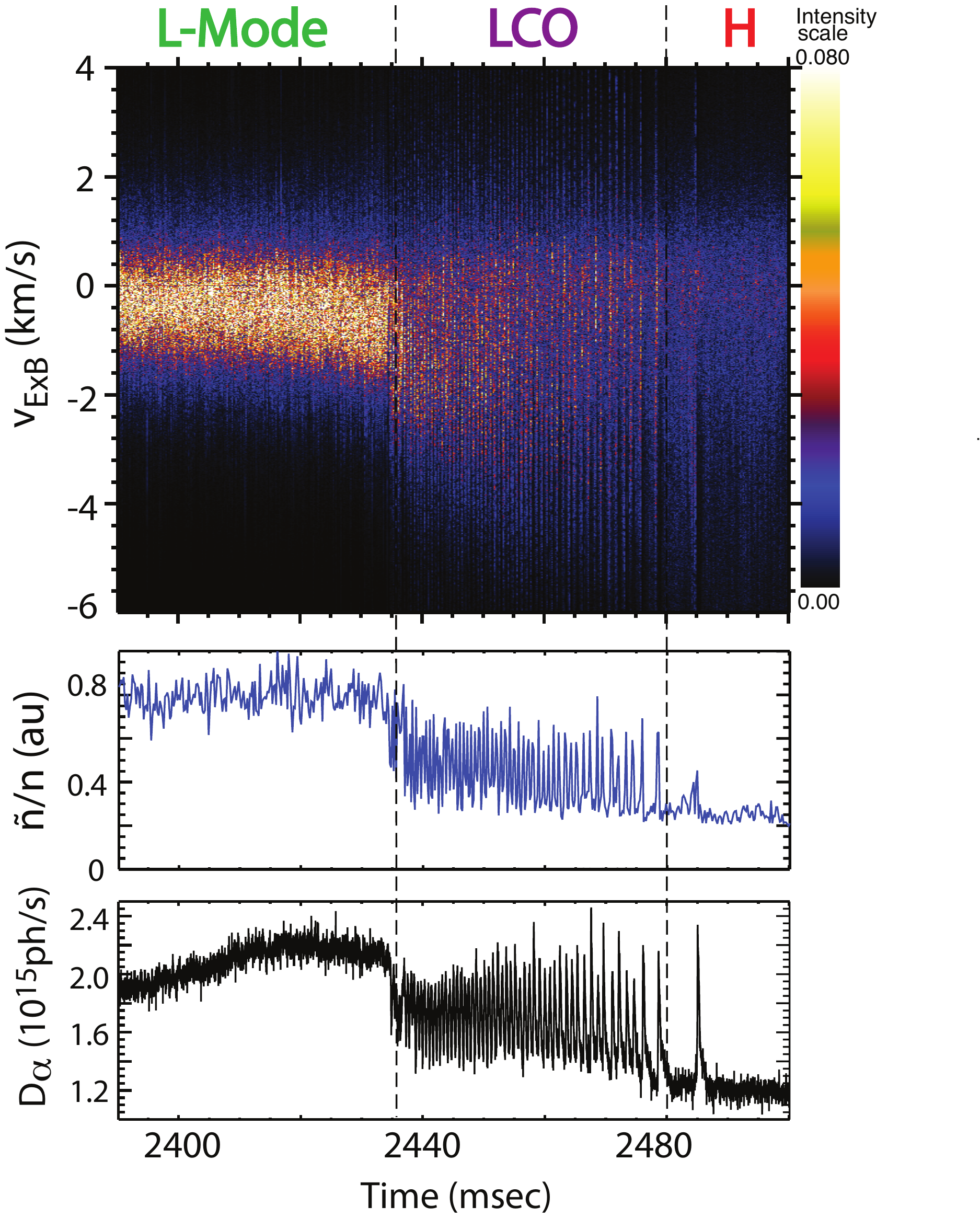}\caption{Time traces of $v_{E\times B}$ shear velocity, density fluctuation amplitude and $D_{\alpha}$ recycling light emission at $\rho=0.96$ during an L-H transition. This work studies the L-mode in the near-edge, defined here as $0.80\leq\rho \leq0.96$, and focuses at $t=2400$ ms, before transition times $t_{\rm LCO}=2436$~ms and $t_{\rm H}= 2480$~ms shown by vertical dashes.} \label{LotharDIIID}\end{figure}

The subject of this study is DIII-D discharge $\#$153624. This discharge exhibits three L-H transitions, where the first was induced with Electron Cyclotron Resonance Heating (ECRH) and the following two with Neutral Beam Injection (NBI) heating. Figure \ref{LotharDIIID} shows highly resolved time-traces taken at $\rho=0.96$ during the second L-H transition (since we focus our study on $t_0=2400$ ms, only data relevant to this transition is shown). 
Note the remarkable change in density fluctuation amplitude $\tilde{n}/n$ and divertor $D_{\alpha}$ recycling light emission as the plasma changes its operational state from L-mode to H-mode via an extended phase of Limit Cycle Oscillations (LCOs)\cite{Lothar, Schmitz17}. The $\bm{E}\times \bm{B}$ velocity $v_{E\times B}$ and density fluctuation amplitude were obtained with the Doppler Backscattering (DBS) diagnostic \cite{Hillesheim10}, which probes a wavenumber range of $0.3 \lesssim k_y \rho_s \lesssim 0.6$. 

The plasma is in the lower single-null shape, where the ion $\nabla B$-drift direction is towards the single active X-point and the power threshold for the L-H transition is relatively low (as compared to the upper single-null shape) \cite{Wagner}. Therefore, the neutral beam is deliberately operated at a relatively low beam power of $1.1$~MW, which is in the vicinity of this power threshold. In steady-state operation, this heating power plus Ohmic heating is comparable to the total heat flux relevant for flux-matching gyrokinetic simulations. Note that the heat flux is typically carried by radiation losses and by radial conductive and convective transport due to microturbulence in the electron and ion channels, which are sometimes difficult to separate empirically at moderate to high collisionality. 

This work will focus on a constant time $t_0=2400$ ms, because this is when the turbulence is still in a quasi-steady equilibrium state before the L-H transition begins (see Fig.~\ref{LotharDIIID}). At $t_0$ the plasma has temperature and density profiles as shown below in Figure \ref{neplot}.
\begin{figure}[!htb] \includegraphics[width=0.457\textwidth]{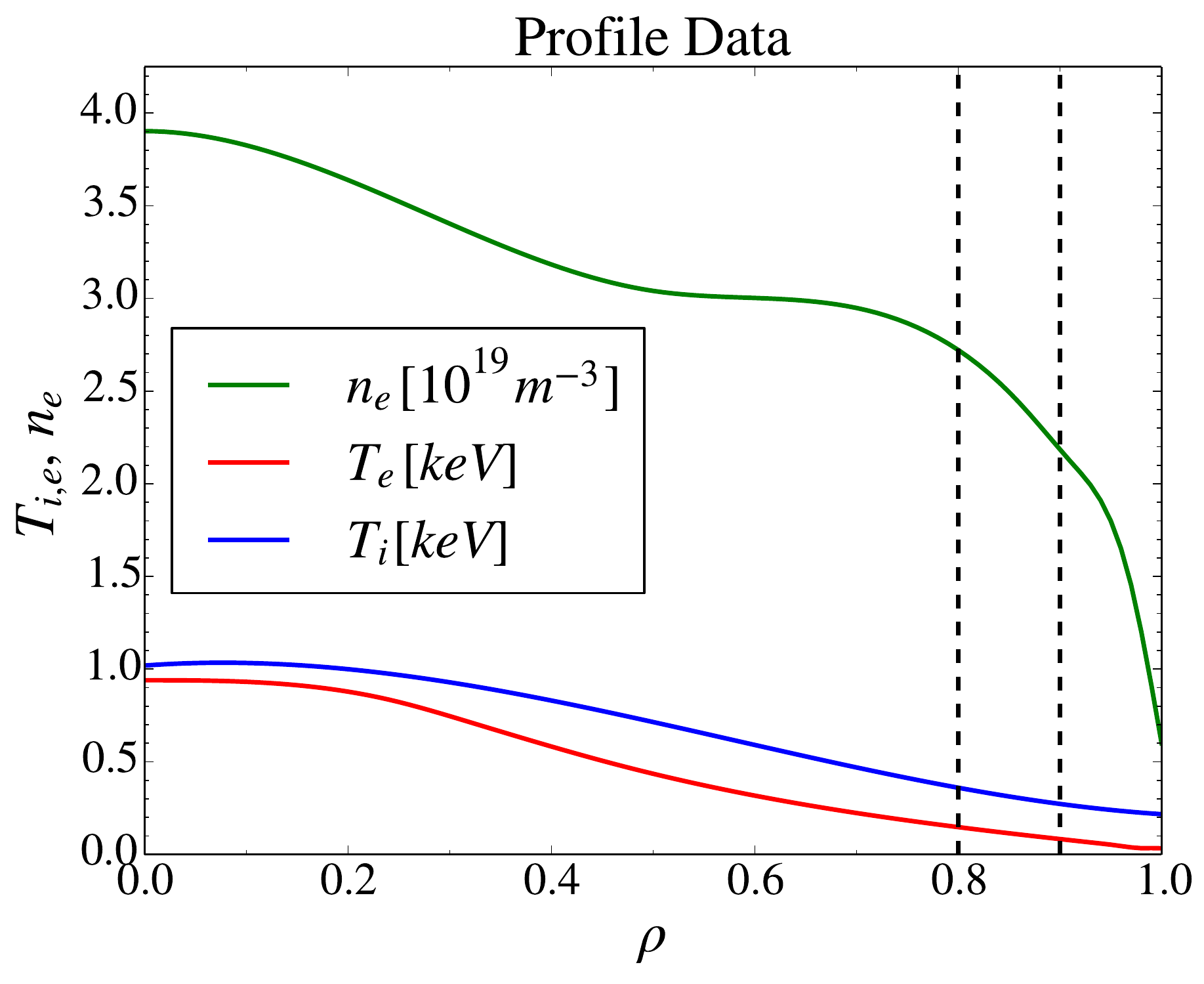} \caption{Experimental profiles of electron density (\emph{green}), electron temperature (\emph{red}) and ion temperature (\emph{blue}). This work studies the L-mode plasma in the near-edge with nonlinear simulations at $\rho=~0.80$ and $\rho=0.90$ as indicated by the vertical dashes.}  \label{neplot}\end{figure}
Both the electron temperature and density are measured by the Thomson scattering diagnostic. This method produces highly resolved profile data for the electrons, but cannot be used effectively for the ions. This is due to their much lower Thomson scattering cross section $\sigma_{t}$, which scales as $\sigma_{t}\propto 1/ m_j^2$, where $m_j$ is the mass of the scattered charge. Therefore, a Charge Exchange Recombination Spectroscopy (CER) diagnostic is used for the impurity ions, which studies the emission lines from neutral beam injection. Specifically, the impurity ion CER diagnostic detects predominantly charge exchange recombination radiation from fully ionized impurity carbon ions ($C^{6+}$). The measured carbon temperature profile is assumed to be equal to the deuterium ion temperature profile in this discharge.

\begin{figure}[!htb] \includegraphics[width=0.457\textwidth]{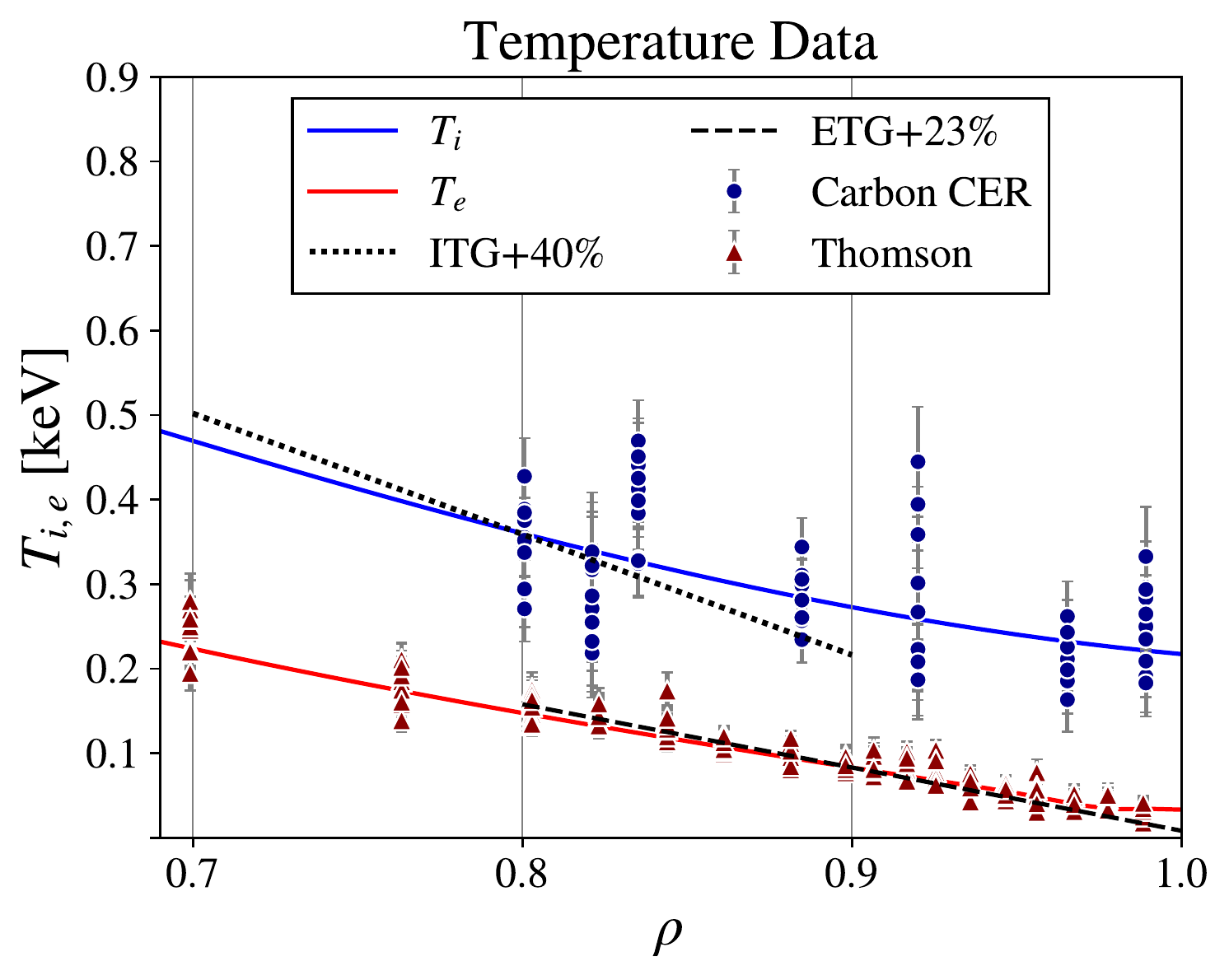}\caption{Close-up view of electron (\emph{red line}) and ion temperature profiles (\emph{blue line}) with experimental data from the CER (\emph{blue circles}) and Thomson diagnostics (\emph{red triangles}). The data was collected in a $\pm20$ ms time window centered at $t_0=2400$~ms. The total (systematic and statistical) uncertainty $\sigma$ in the ion temperature gradient~(ITG) at $\rho=0.80$ is estimated to be $\sigma_{\rm ITG}\sim 25\%$. The uncertainty in electron temperature gradient (ETG) at $\rho=0.90$ is estimated to be $\sigma_{\rm ETG}\sim 18\%$ (see Appendix). For reference, we show a $40\%$ increase in ITG at $\rho=0.80$ (\emph{dotted line}) and a $23\%$ increase in ETG at $\rho=0.90$ (\emph{dashed line}).} \label{TiProf}\end{figure}

In general, the comparatively lower detection number statistics of CER data versus Thomson data causes the statistical uncertainty of the ion temperature to be larger than the uncertainty in the electron temperature (see Fig.~\ref{TiProf}). For instance, the statistical uncertainty for the ion temperature gradients at $\rho=0.80$ is estimated to be $\sigma_{\text{ITG, stat}}\sim~15\%$ (see Appendix). On the other hand, the statistical uncertainty in the electron temperature gradient (ETG) at $\rho=0.90$ is estimated to be $\sigma_{\text{ETG, stat}}\sim 8\%$. Moreover, the systematic uncertainty for both electron and ion temperature gradients is estimated to be $\sigma_{\text{sys}}\sim 10\%$. Recall that only slightly more than two thirds ($68\%$) of normally distributed measurements fall within $1\sigma_{\text{stat}}$, while most ($95\%$) fall within $2\sigma_{\text{stat}}$ and nearly all ($99.7\%$) fall within $3\sigma_{\text{stat}}$ of the mean. Therefore, changes in temperature gradient of up to $\leq2\sigma_{\text{stat}}+1\sigma_{\text{sys}}$ can reasonably be attributed to combined uncertainties in the measurement and model assumptions. 

The physical parameters at four radial positions in the near-edge and edge region are summarized in Table \ref{params}. 
\begin{table}[!ht]
\begin{tabular*}{\columnwidth}{c @{\extracolsep{\fill}} ccccc}
  \hline \hline
  $\rho$ & 0.80 &0.85& 0.90 & 0.95  \\  \hline
  Time/[s] & 2.40& 2.40& 2.40 &2.40\\
  $\hat{s}$ &  1.41&1.98 & 2.98 & 5.18  \\ 
  $q$ &  2.86&3.15 &3.69 &4.47 \\
  $\omega_{T_i}$ &  2.78 & 2.80 & 2.68 & 2.32 \\ 
  $\omega_{T_e}$ & 4.69 &5.62 &7.32  & 13.51\\ 
  $\omega_{n_e}$ & 1.34 & 2.21& 2.91 & 7.05 \\ 
  $\beta_e\left/[\%] \right.$ & 0.0557& 0.0396& 0.0252 & 0.0132\\ 
  $T_i\left/[\rm keV]\right.$ &  0.360 &0.320 &0.281& 0.244\\ 
  $T_e\left/[\rm keV]\right.$ & 0.148 & 0.119 &0.0831&0.0531\\ 
  $n_e\left/[10^{19} \, {\rm m}^{-3}]\right.$ & 2.72 &2.68 &2.19 &2.12\\ 
  $Z_{\rm eff}$ &  1.80 &  1.80 &1.80 & 1.80\\ 
  $\nu_{ei}\left/[c_s/L_{\rm{ref}}]\right.$ & 7.28 &  10.9&17.8 & 35.21\\ 
  $B_{\rm ref}\left/[{\rm T}]\right.$ &  1.70 &1.70  &1.70& 1.70\\ 
  $L_{\rm ref}\left/ [{\rm m}]\right.$ &  0.770 & 0.770 &0.770&0.770\\ 
  $\rho_{s}\left/ [10^{-3}{\rm m}]\right.$&  1.03& 0.927 &0.775&0.619\\
  $n\left/[k_y \rho_s]\right.$ & 208 & 224 & 242 & 264 \\
  \hline \hline
\end{tabular*}
\caption{Physical parameters for radial positions in the near-edge region, with variables as defined in the text.}
\label{params}
\end{table}
The variables in the table are defined as follows. The logarithmic gradients are defined as
\begin{equation}
\omega_{X}=-\frac{1}{X} \frac{dX}{d\rho}\,, \text{ with } X\in \{T_i, T_e, n\}\,,
\end{equation}
where $\rho=(\Phi/\Phi_{\rm edge})^{1/2}$ is the toroidal flux radius normalized by $\Phi_{\rm edge}$, which is the toroidal flux at the separatrix. The shear parameter is given by
\begin{equation}
\hat{s}  = \frac{\rho}{q}\frac{d q}{{d} \rho}\,,
\end{equation}
where $q$ is the safety factor. The electron beta is defined as the ratio of thermal to magnetic pressure, 
\begin{equation}
\beta_e=\frac{n_e T_e}{B_{\rm ref}^2/2 \mu_0 }\,,
\end{equation}
where $B_{\rm ref}$ is the magnetic field on axis and the other variables take their usual meaning. The effective atomic number of the plasma, $Z_{\rm eff}={\Sigma_i Z_i^2 n_i}/{n_e}$, is greater than that of a pure deuterium plasma ($Z=1$) mostly due to carbon impurities that enter the plasma from the divertor and the wall. The electron-ion collision frequency is defined as
\begin{equation}
\nu_{ei}= \frac{Z_{\rm eff} n_e e^4 \ln{\Lambda}}{2^{7/2} \pi \epsilon_0^2 m_e^{1/2}T_e^{3/2}}\,,
\end{equation}
where $\ln \Lambda$ is the Coulomb logarithm and other variables take their usual meaning. Convenient reference lengths are given by $L_{\rm ref} =~\sqrt{\Phi_{\rm edge}/\pi B_{\rm ref}}$ and $\rho_s=\sqrt{T_e m_i /e^2 B_{\rm ref}^2}$. For example, we will use $\rho_s$ to rescale the wavenumber of turbulent modes as $k_y \rho_s$. For a better comparison with other work, a conversion from $k_y \rho_s$ to the toroidal mode number $n$ is useful. The conversion factor is given by the relation \cite{DTold13}
\begin{equation}
\frac{n}{k_y \rho_s}= \frac{\rho}{q} \frac{L_{\rm ref}}{\rho_s}\,,
\end{equation}
which is evaluated numerically in the bottom row of Table \ref{params}.

\begin{figure}[!htb] \includegraphics[width=0.457\textwidth]{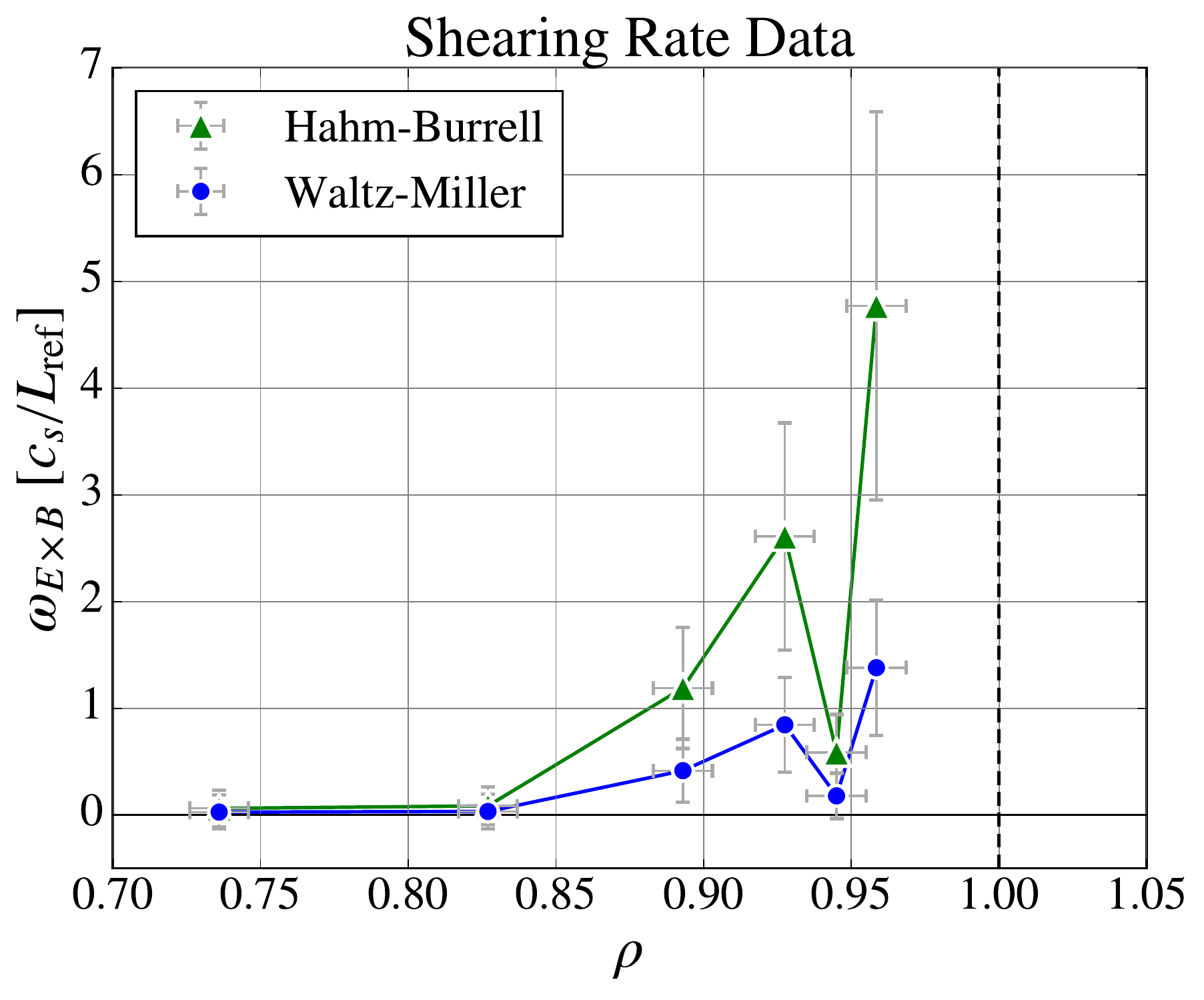} \caption{The Hahm-Burrell shearing rate in the edge region, as evaluated by magnetic field geometry and the Doppler Backscattering measurements at the outboard midplane (\emph{green triangles}). A flux-surface average of the Hahm-Burrell rate is known as the Waltz-Miller rate and commonly used in the simulation community (\emph{blue circles}). The flow shear generally increases towards the separatrix~(\emph{dashed line}).} \label{WexB}\end{figure}

In this work, we also take into account the effect of the $\bm{E}\times \bm{B}$ shearing rate, which is a function of both the radial electric field gradient and the magnetic field geometry. For realistic geometry, the $\bm{E}\times \bm{B}$ shearing rate is given by the so-called Hahm-Burrell formalism \cite{HahmBurrell95}, 
\begin{equation}
\omega_{E\times B}= \frac{c R B_{\theta}}{B} \frac{d}{dr}\left( \frac{E_r}{B_{\theta} R} \right) \,,
\end{equation}
where $E_r$ is the radial electric field, $B_{\theta}$ is the poloidal component of the magnetic field $B$, and $R$ is the major radius. Note that $\left({E_r}/{B_{\theta} R}\right)$ is constant on the flux surface, but the factor ${c R B_{\theta}}/{B}$ is not constant (it is larger on the outboard midplane). In the simulation community, a flux-surface-averaged formalism for the $\bm{E}\times \bm{B}$ shearing rate in shaped geometry is often used for simplicity~\cite{WaltzMiller99},
\begin{equation}
\omega_{E\times B}\approx \frac{r}{q} \frac{d}{dr}\left( \frac{E_r}{B_{\theta} R} \right)\,,
\end{equation}
where the factor preceding the derivative, $r/q$, is now a flux function. The present work will also employ this so-called Waltz-Miller formalism. The experimental shearing rate values are inferred using data from the magnetic field geometry and the Doppler Backscattering diagnostic (see Fig.~\ref{WexB}). 

\section{Gyrokinetic Simulation Method \label{meth}}
Throughout this work, we employ the gyrokinetic turbulence code GENE \cite{GENE}. 
The gyrokinetic approximation reduces the six-dimensional phase space to five dimensions by averaging over a charged particle's gyro-motion perpendicular to the magnetic field, and removes several phenomena on small space-time scales. The resulting Vlasov equation can be coupled self-consistently to Maxwell's equations. GENE is a Eulerian code that solves the relevant equations on a field-aligned coordinate system ($\bm{x}$, $v_{\parallel}$, $\mu$), which minimizes the necessary number of grid points \cite{Beer95, Xanthopoulos06}. Here, $v_{\parallel}$ represents the velocity along the field lines and $\mu$ is the magnetic moment resulting from the gyro-averaged motion of a charge. 

GENE can model the plasma in a local (toroidal flux-tube\cite{Cowley1991}) or a global (radial annulus) simulation domain. The local approximation is preferred where applicable, because periodic boundary conditions in both the radial ($x$) and binormal ($y$) directions invite numerically efficient spectral methods. 
The local approximation holds where the turbulent correlation lengths are smaller than the gradient scale lengths, such that the plasma parameters do not vary much across a typical turbulent structure. This condition appears to be satisfied in the near-edge region of the L-mode plasma considered here, so we will use the local approach throughout this work.

In our nonlinear simulations we consider two particle species (electrons and Deuterons), and thereby neglect the effect of carbon as motivated by linear simulations presented below in section \ref{linsim08} and \ref{edge}. Electromagnetic effects are included in our simulations by solving for the parallel component of Amp\`ere's law. Moreover, GyroLES techniques were used to model dissipation at unresolved wavenumbers, thus avoiding the unphysical build-up of energy at the highest resolved wavenumbers \cite{GyroLES11, GyroLES12, Banon14}. The flux-tube geometry is calculated with the TRACER-EFIT interface \cite{Xanthopoulos09}. Due to the high collisionality in the near-edge region, a collision operator developed by Sugama \emph{et al.} was used \cite{Sugama09}.  

Closer to the edge region, at $\rho=0.90$, it becomes evident that the effects of sheared flow on turbulent heat transport need to be included, while the shearing rate at $\rho=0.80$ is negligible (see Fig.~\ref{WexB}). The radial flow shear can have an important effect, because it shears turbulent eddies in the poloidal direction, which increases their poloidal correlation length and reduces their radial correlation length \cite{Biglari90, Burell99, Shesterikov13}. This can lead to experimentally relevant improvements in particle and energy confinement. Specifically in GENE, the constant Waltz-Miller shearing rate throughout the flux-tube is implemented using a method developed by Hammett~\emph{et~al.}\cite{HammettExB06}. Here, a transformation into the co-moving coordinate system of the equilibrium flow and a discrete time evolution of the sheared radial wavenumber greatly reduce computational intensity while maintaining acceptable numerical accuracy~\cite{HammettExB06}.

In order to investigate the interaction between strong ETG streamers and ion-scale modes, it is instructive to carry out multi-scale simulations resolving both ion and electron scales. These simulations are very computationally intensive and cannot currently be carried out resolving the full wavenumber domain of linearly unstable modes. Therefore, a reasonable reduction in the resolved wavenumber domain is sought with nonlinear single-scale simulations. For example, nonlinear electron-scale simulations with varying $k_{y, \rm{max}} \rho_s$ can help determine the reliability of GyroLES techniques; they help identify a reasonable maximum extent of the wavenumber domain that still captures the main nonlinear turbulent transport. A similar nonlinear scan in $k_{y, \rm{min}} \rho_s$ is carried out at the ion scales to determine a feasible multi-scale simulation domain that still captures the majority of the ion-scale physics. With this method, we are able to carry out multi-scale simulations with realistic electron-Deuteron mass ratio and realistic geometry for the first time in the near-edge, at $\rho=0.80$.

Throughout this work, the simulation domain in velocity space extends in the parallel direction up to $v_{\parallel,\rm max}=3 v_{\text{th},j}$, where $v_{\text{th},j}= \sqrt{2T_{0,j}/m_j}$ is the thermal velocity. In the perpendicular direction, it extends to $\mu_j = 9 T_{0,j}/B_{\rm ref}$. Further details of the simulation method, such as the number of grid points or the radial size of the simulation box, are described together with the results of the relevant nonlinear simulation. Throughout this work, the GyroLES methods, the Sugama collision operator and the inclusion of $\bm{E} \times \bm{B}$ shear effects mentioned above are of particular relevance.

\section{Results at first radial position ($\rho=\mathbf{0.80}$)\label{linsim08}}
All microinstabilities ultimately saturate due to nonlinear interactions between modes of differing wavenumbers. Nonetheless, linear simulations often give useful insight into the nature of the nonlinear instabilities. Generally, the distribution of linear growth rates in wavenumber space highlights the scales of the nonlinear turbulent drive. This can inform the size of the nonlinear simulation box. Moreover, sensitivity of these linear growth rates to changes in physical parameters such as temperature gradients can help identify the nature of modes encountered in nonlinear simulations. Therefore, we will first present linear simulation results, followed by nonlinear simulation results.

\subsection{Linear Simulations \label{gammarule}}
\begin{figure}[!htb] \includegraphics[width=0.457\textwidth]{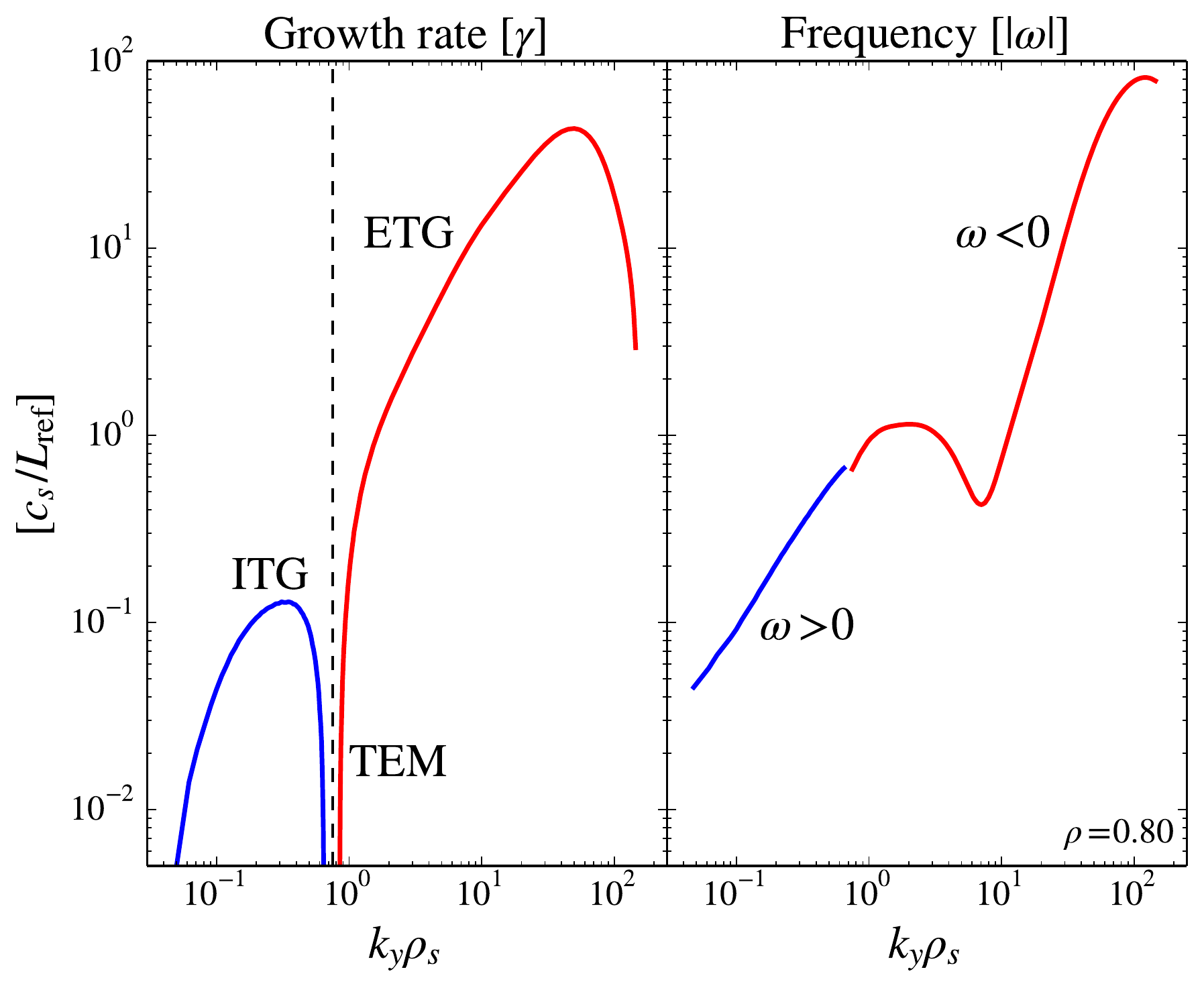}\caption{Linear growth rates $\gamma$ (\emph{left}) and absolute values of frequency $|\omega|$ (\emph{right}) as a function of poloidal wavenumber $k_y \rho_s$. The growth rates are separated by a stable region (where $\gamma<0$) at $0.68<k_y \rho_s < 0.90$. For nonlinear simulations, we define the ion scales in the domain $0.05\leq k_y \rho_s \leq 0.80$ and the electron scales in the somewhat overlapping domain $0.70\leq k_y \rho_s \leq 180$.}
\label{lineare}\end{figure}

We use both the initial value solver and the eigenvalue solver in GENE to find linear growth rates of modes at the electron and ion scales. 
The resolution for all linear simulations in this work, which analyze each $k_y$ mode individually, is $\left(n_x, n_z, n_{v_{\parallel}}, n_{\mu}\right)=\left(31, 32, 32, 24\right)$. At $\rho=0.80$, linearly unstable ion temperature gradient (ITG) modes can be identified by the positive frequencies associated with the diamagnetic drift direction of the ions. Similarly, trapped electron modes (TEM) and electron temperature gradient (ETG) modes are identified by the negative frequencies associated with the electron diamagnetic drift direction (see Fig.~\ref{lineare}). The frequencies of these modes are much smaller than the ion-cyclotron frequency such that the gyrokinetic approximation can be used to describe these modes. The unstable ITG and TEM/ETG modes are separated by a stable region in wavenumber space ($0.68< k_y \rho_s <0.90$). These separate domains allow us to clearly define ion scales ($0.05 \leq k_y \rho_s \leq 0.80$) and somewhat overlapping electron scales ($0.70\leq k_y \rho_s \leq180$) for separate nonlinear analysis. 

\begin{figure}[!htb] \includegraphics[width=0.457\textwidth]{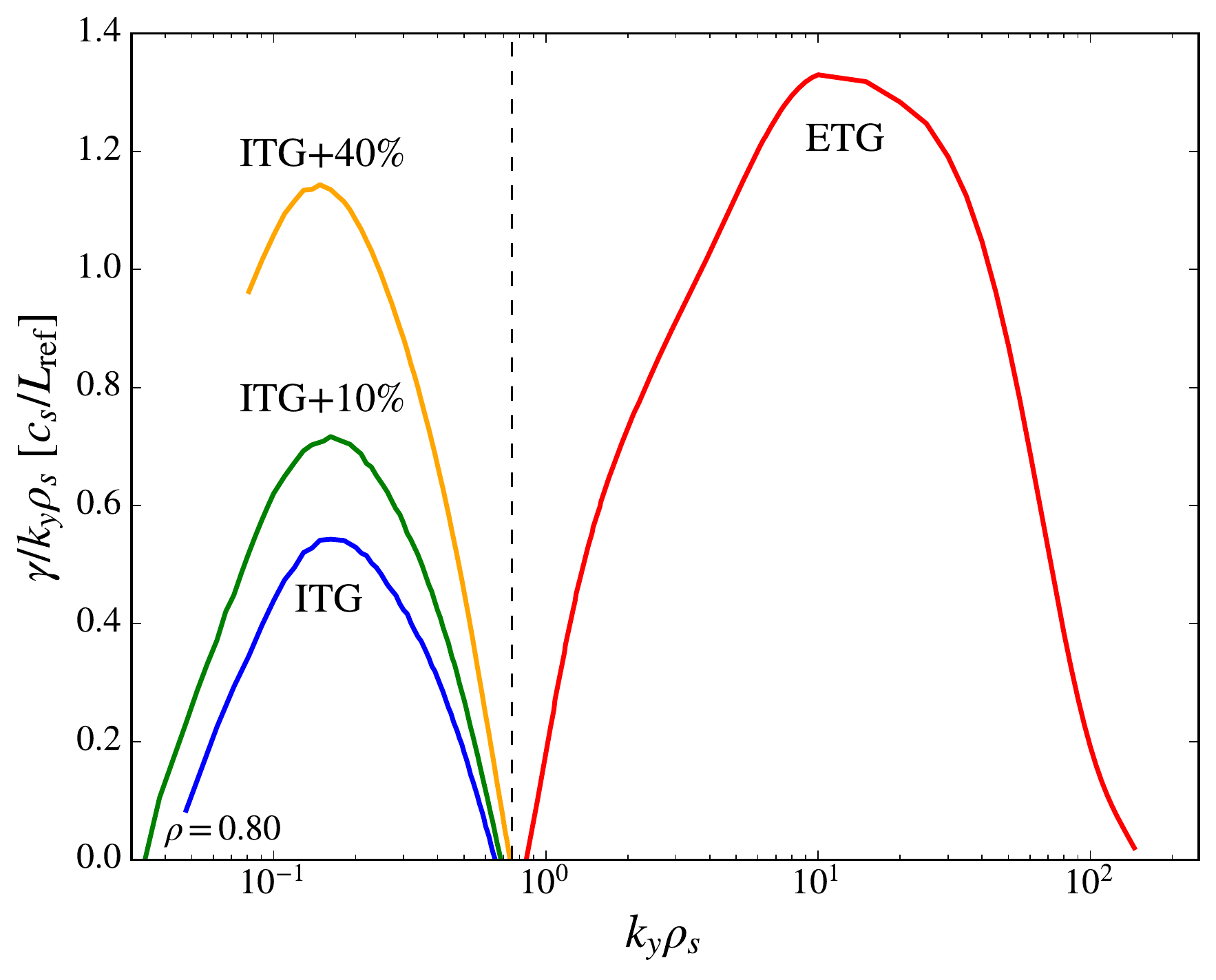}\caption{The ratio of linear growth rates and poloidal wavenumber ($\gamma/k_y\rho_s$) as a function of wavenumber. For cases where the ITG is nominal (\emph{blue}) or increased by $10\%$ (\emph{green}), the peak of said ratio is smaller at the ion scales than at the electron scales. For the case where the ITG is increased by $40\%$ (\emph{orange}), the difference between the peaks is much less pronounced.}
\label{Staebler}\end{figure}

It is not clear how the heat fluxes found with nonlinear ion-scale and electron-scale simulations contribute to the collective heat flux. Generally, multi-scale simulations that simultaneously resolve both scale ranges are necessary to answer this question. These simulations are very expensive and may not be possible in all scenarios. A heuristic rule has emerged from pioneering work~\cite{JenkoJap04, Goerler08PRL, Goerler08PoP, Waltz07} using a reduced mass ratio ($\sqrt{m_i/m_e}= 20$) and $\hat{s}-\alpha$ geometry, with $\alpha=0$. Namely, if the ratio of maximum growth rates at the electron and ion scales is much larger than the square root of the mass ratio,
\begin{equation}
{\gamma_{\rm ETG}^{\rm max}}\left/{\gamma_{\rm ITG}^{\rm max}}\right. \gg \sqrt{{m_i}\left/{m_e}\right.}\,,
\end{equation}
then multi-scale effects could be present. This is because the contributions by the electron-scale turbulence to the overall heat transport could be important. Otherwise, turbulent structures at the ion scales disrupt the efficient heat transport at the electron scales. In our case the mass ratio is $\sqrt{m_i/m_e}\approx 60$ and the ratio between the maximum growth rates is $\gamma_{\rm ETG}^{\rm max}/\gamma_{\rm ITG}^{\rm max} =338$. Therefore our linear simulations with the nominal experimental parameters indicate that multi-scale effects could be present. For an increase in the ion temperature gradient by $\sim40\%$, we get $\gamma_{\rm ETG}^{\rm max}/\gamma_{\rm ITG+40\%}^{\rm max}=161$, so multi-scale effects could also be present at this point in parameter space, according to this rule-of-thumb. More recently, a model for saturation of multi-scale turbulence by zonal flow mixing has been proposed\cite{Staebler17}. In this model, an important parameter is the RMS velocity of zonal flows ($V_{\text{ZF}}$), which saturates at $V_{\text{ZF}}=\text{Max}\left(\gamma/k_y\right)$, where $\gamma$ is the linear growth rate of a turbulent mode with wavenumber $k_y$. According to this model, multi-scale effects could be present when 
\begin{equation}
\text{Max}\left(\gamma_{\text{ETG}}/k_y\right)\geq\text{Max}\left(\gamma_{\text{ITG}}/k_y\right)\,.
\end{equation}
This criterion has been validated on recent multi-scale simulations using realistic mass ratio and geometry\cite{HowardNF13, HowardNF16}. For our linear simulations at $\rho=0.80$, this criterion is satisfied for the cases where the ion temperature gradient (ITG) is at its nominal value or increased by $10\%$; it is also satisfied when the ITG is increased by $40\%$, but by a small margin (see Fig.~\ref{Staebler}). 
Therefore, single-scale simulations are likely not sufficient for the cases considered here and multi-scale simulations will need to be carried out. This will be presented in the following subsection \ref{nonlinsim}. 

\begin{figure}[!htb] \includegraphics[width=0.457\textwidth]{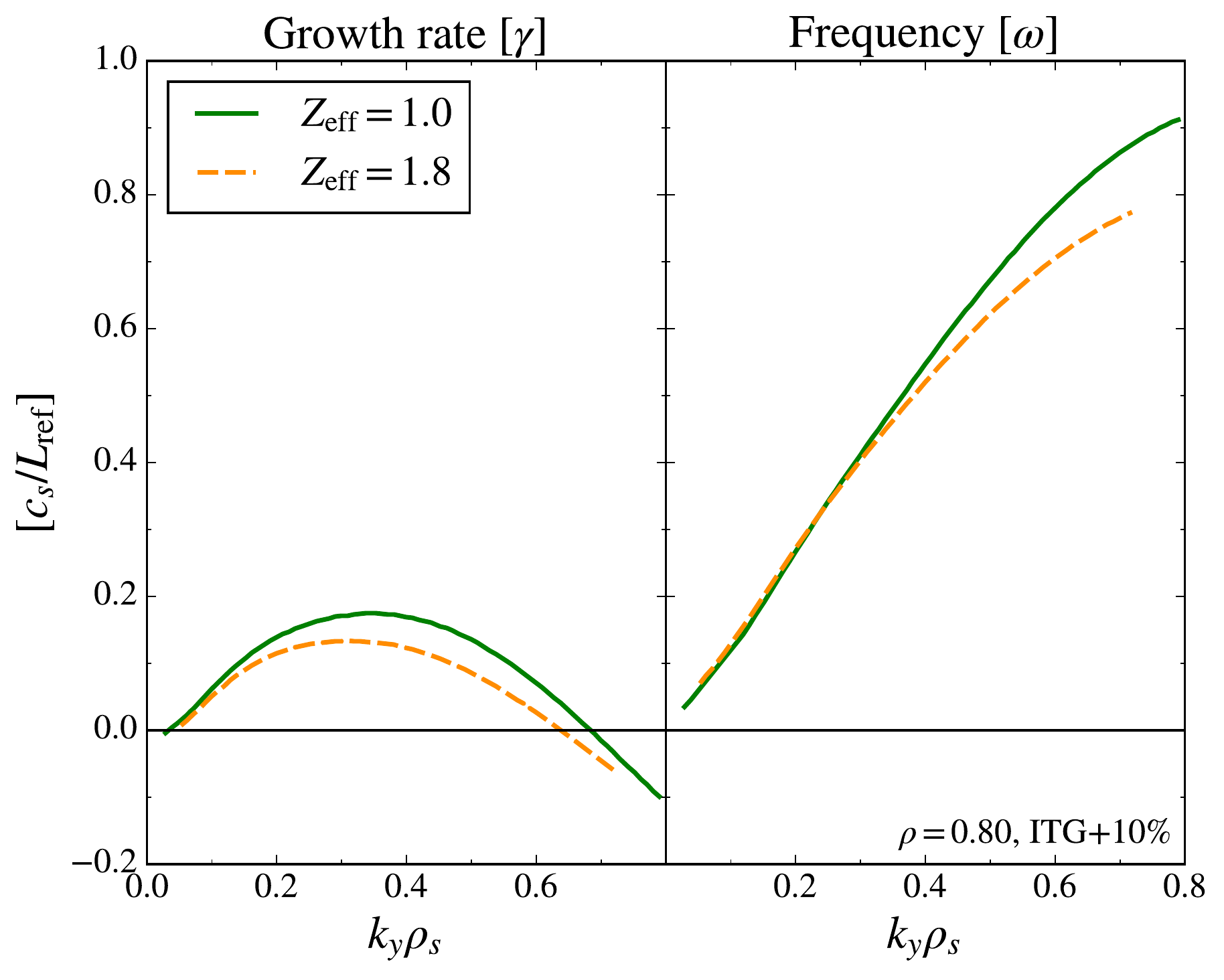}\caption{Linear growth rates and frequencies for a pure deuterium plasma (\emph{green}) and for a plasma with an added carbon impurity species corresponding to $Z_{\rm eff}=1.8$ (\emph{orange}). For both cases, the ITG is increased by $10\%$, which is above the nonlinear critical gradient as shown in Figure \ref{nonlinear}. The carbon impurity has a small effect on the linear growth rates compared to the pure deuterium plasma.}\label{Zeff}\end{figure}

It is not \emph{a priori} clear whether impurities significantly affect turbulence in the deuterium plasma. The DIII-D tokamak has a carbon divertor and wall that add carbon as the main impurity to the deuterium plasma. Using the impurity ion Charge Exchange Recombination (CER) diagnostic\cite{Gohil91}, the plasma is observed to have an effective atomic number of $Z_{\rm eff}=1.80$ (see Table \ref{params}). In order to quantify the effect of this impurity, linear simulations are carried out with three particle species, namely deuterium and carbon ions, and electrons. We find that the carbon impurity has a negligible effect on the linear growth rates of ITG modes (see Fig.~\ref{Zeff}). 
Due to this observation, to first order in accuracy, carbon impurities can be neglected in our nonlinear simulations at $\rho=0.80$.
In summary, linear simulations at $\rho=0.80$ identify coexisting ITG and ETG modes that could engage in multi-scale interactions.

\subsection{Nonlinear Ion-scale Simulations \label{nonlinsim}}
Fully nonlinear gyrokinetic simulations can be used to diagnose experiments in the hope to improve them in the future. When carrying out these simulations, several steps have to be taken to accurately extract the radial heat flux of the system. After each simulation, we ensure that the perpendicular box size $(L_x, L_y)$ accommodates several correlation lengths of the turbulence. This reduces the effect of boundary conditions on the turbulent structures and avoids their end-to-end connection across the boundaries. We check the grid resolution for convergence, $(n_x, n_y, n_z, n_{v_{\parallel}}, n_{\mu})$, by repeating a certain run with higher resolution in certain dimensions and checking for consistency with previous runs. This is particularly important closer to the edge, where high shear ($\hat{s}>2$) demands high radial resolution \cite{DTold13}. Computational expense on the order of ten million central processing unit (CPU) hours (MCPUh) of multi-scale simulations presently restricts convergence tests to the single-scale simulation domain. To ensure an accurate reading of the simulated heat flux, it is averaged over a time greatly exceeding the turbulent correlation time.

\begin{figure}[!htb] \includegraphics[width=0.457\textwidth]{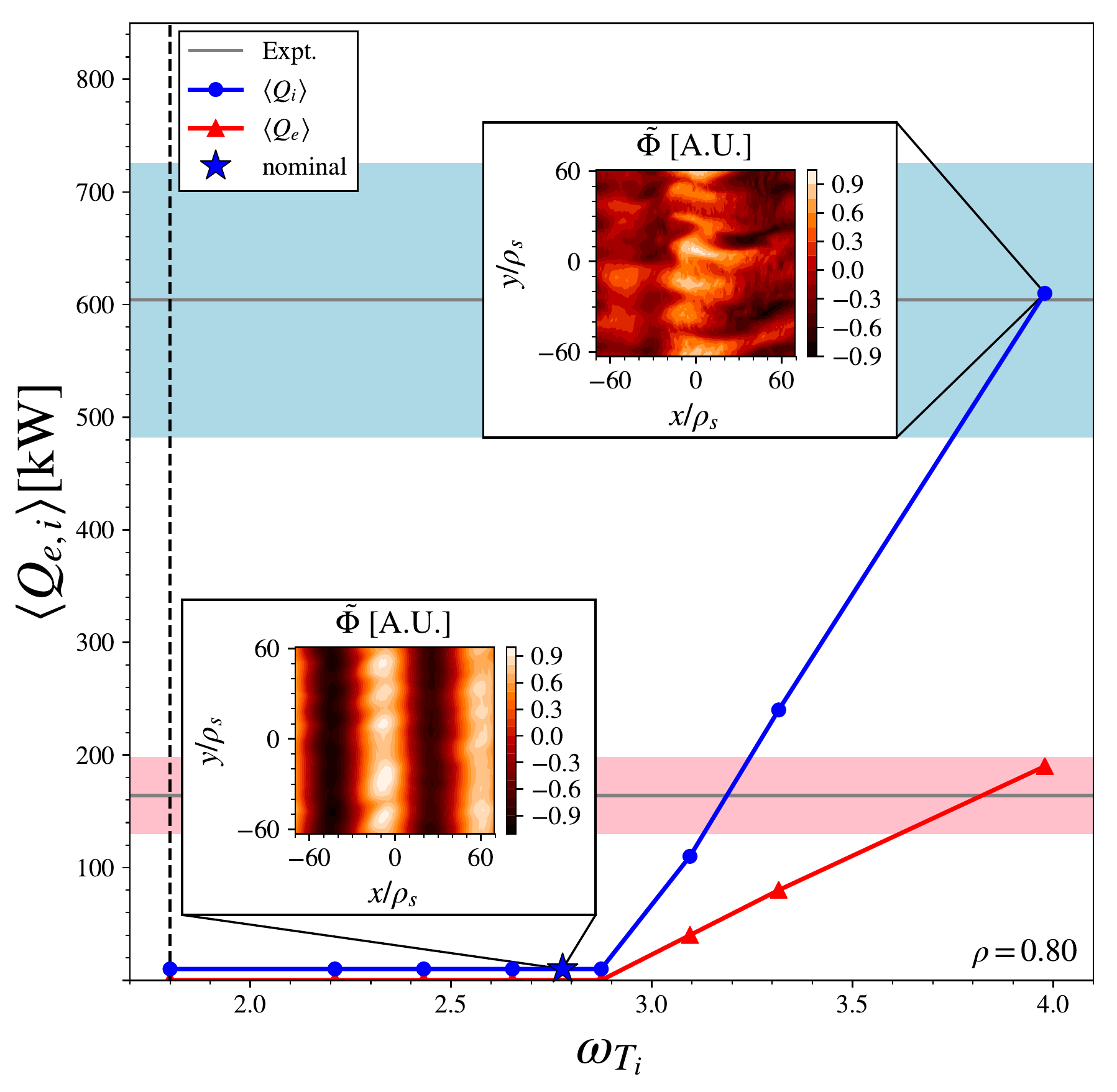}\caption{Nonlinear ion-scale simulations with a clear Dimits shift. The vertical black dashed line marks the linear critical ion temperature gradient $\left(\omega_{T_i}\right)_{\mathrm{crit}}=1.80$. A strong poloidal zonal flow is found at nominal parameters (\emph{lower inset}) and large-scale turbulence is found in the form of radially-elongated streamers at higher gradients (\emph{upper inset}). The uncertainty in the experimental heat fluxes is $\pm 20\%$ (\emph{shaded regions}). Increasing the ITG by $\sim40\%$ recovers the experimentally inferred heat fluxes of, remarkably, both the ions and electrons.} \label{nonlinear}\end{figure}

Nonlinear ion-scale simulations are performed with $(n_x, n_y, n_z, n_{v_{\parallel}}, n_{\mu})=(256, 64, 24, 32, 24)$ grid points and perpendicular box size $(L_x, L_y)=(140 \rho_s, 126\rho_s)$. For the nominal experimental parameters as input we see a nonlinear quench of radial ion heat fluxes and the formation of a strong poloidal zonal flow (see lower inset, Fig.~\ref{nonlinear}). 
Continuing the simulation for several hundred time units ($\sim 400$  $L_{\rm ref}/c_s$) to ensure nonlinear saturation, we find a time average ion heat flux that indicates nonlinear ``stability'' of ITG modes. In this case, the primary ITG instability leads to a secondary instability that generates a poloidal zonal flow and quenches the radial ion heat transport to $\langle Q_i \rangle \sim 10$~kW (see lower inset of Fig.~\ref{nonlinear}). Increasing the ion temperature gradient by $\sim 3.5\%$ to $\omega_{T_i}\simeq2.9$ marks the onset of nonlinear instability of ITG modes.
This indicates a strong Dimits shift \cite{Dimits}, defined as the difference between the nonlinear and linear critical temperature gradients for onset of turbulent transport.
Further increasing the ion temperature gradient leads to approximately linear increases in the electron and ion heat fluxes. These heat fluxes are carried by radially elongated turbulent structures (streamers) at the ion scales \cite{Drake88, Cowley1991} (see upper inset of Fig.~\ref{nonlinear}). This behavior is typical of plasmas in the core and has been used to infer the ion temperature gradient with gyrokinetic simulations\cite{Nunami18}.

With the ion temperature gradient increased by $40\%$, our simulations recover the experimentally inferred heat fluxes for, remarkably, both the ion and electron heat channels. This ITG+40\% scenario is consistent with the combined uncertainty of the ion temperature data at the $1.6\sigma$ level, where we write the combined uncertainty as $\sigma=\sigma_{\text{stat}}+\sigma_{\text{sys}}$ (see Appendix). Specifically, the simulations give $\langle Q_i \rangle = 610$ kW and $\langle Q_e \rangle =190 $ kW, while the experimental values obtained with the ONETWO transport code are $\langle Q_i \rangle = (600\pm120)$ kW and $\langle Q_e \rangle = (164\pm33)$ kW (see Fig.~\ref{nonlinear}). 

\subsection{Nonlinear Electron-scale Simulations}
The nonlinear electron-scale simulations are performed with $(n_x, n_y, n_z, n_{v_{\parallel}}, n_{\mu})=(64, 512, 16, 32, 9)$ grid points and perpendicular box size $(L_x, L_y)=(9 \rho_s, 9\rho_s)$. Note that the box size is smaller relative to the ion-scale simulations because the electron-scale domain is defined by linear simulations as $0.70 \leq k_y \rho_s \leq 180$. While finite Larmor radius (FLR) effects for $k_y \rho_s \gtrsim 60$ can be expected to significantly damp ETG modes, note that these high-k modes are mapped to smaller physical wavenumbers by geometric effects (encapsulated in the metric tensor). Thus, high-k modes can contribute meaningfully to the turbulent drive and it is advisable to extend the nonlinear electron-scale simulation domain over the entire wavenumber range of linearly unstable TEM/ETG modes. Note that for our electron-scale simulations we include fully kinetic ions (rather than only adiabatic ions).

\begin{figure}[!htb] \includegraphics[width=0.457\textwidth]{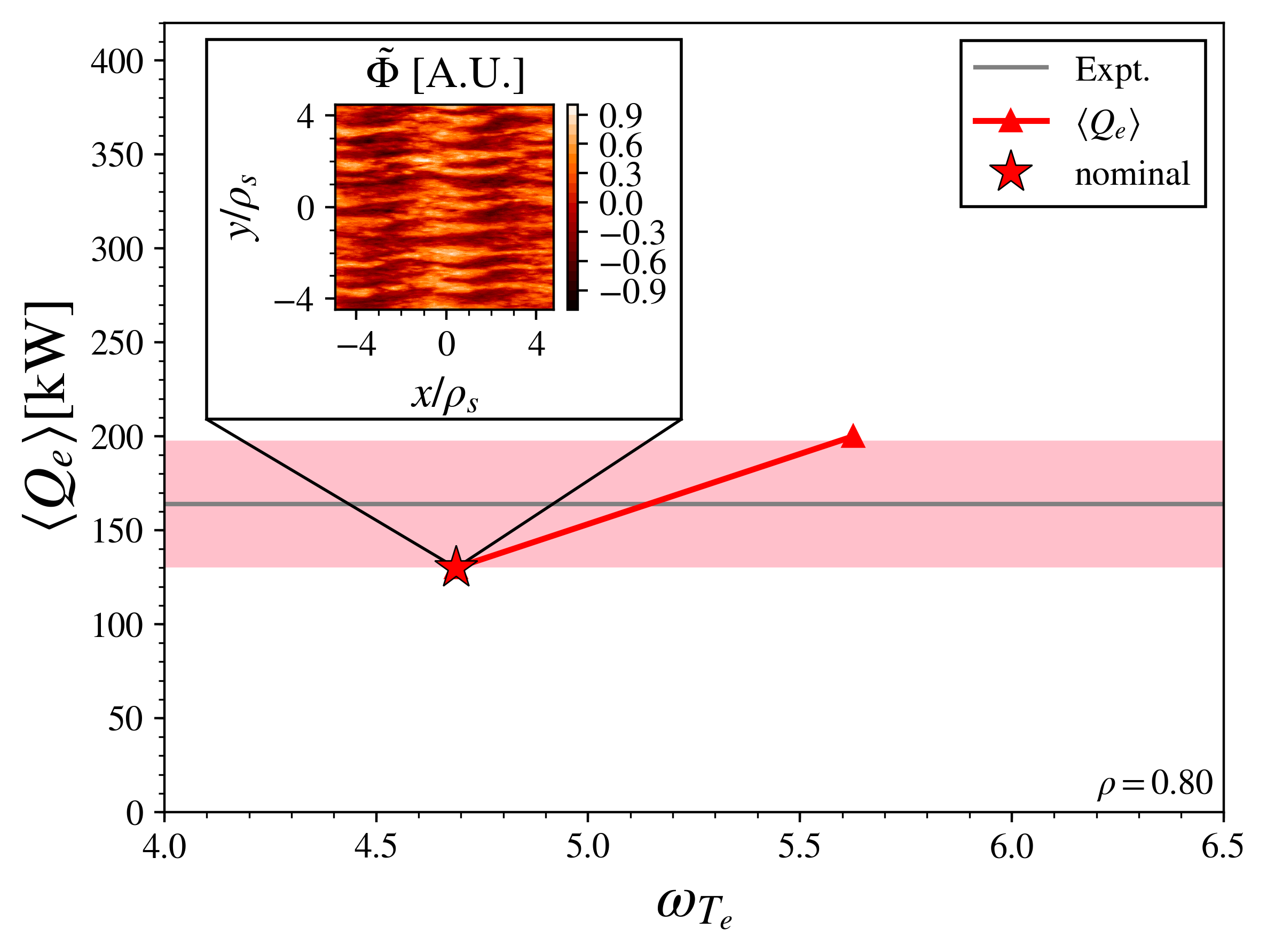}\caption{Dependence of the electron heat flux on the electron temperature gradient for nonlinear electron-scale simulations. Experimentally relevant heat transport could be carried at relatively large wavenumbers~($k_y\rho_s\sim8$) by streamers (\emph{see inset}), which motivates multi-scale simulations.} \label{nonlinearQe}\end{figure}

For the nominal experimental input parameters we find a time average electron heat flux of $\langle Q_e \rangle = 130$ kW. Increasing the electron temperature gradient, which is the main driver of the electron heat flux, by its experimental error of $\sim20\%$, we get a flux of $\langle Q_e \rangle = 200$ kW. This heat flux is within the experimentally inferred range of electron heat flux values obtained with the ONETWO transport code, namely $\langle Q_e \rangle = 164 \pm 33$ kW (see Fig.~\ref{nonlinearQe}). This suggests that electron-scale heat transport could contribute to the overall heat transport, which will need to be studied with multi-scale simulations. Physically, we find that the electron heat flux is carried by radially elongated structures called streamers\cite{JDHPoP01}. These structures are well-defined in a contour plot of electrostatic potential fluctuations, $\tilde{\Phi}(x, y)$ (see inset in Fig.~\ref{nonlinearQe}).

\subsection{Nonlinear Multi-scale Simulations \label{multiscale}}
Linear and nonlinear simulations have indicated that multi-scale interactions could be present. We have therefore carried out the first nonlinear gyrokinetic multi-scale simulations using a realistic mass ratio and experimental input parameters in the near-edge, at $\rho=0.80$. These used on the order of $23$ k processors and $10$ MCPUh on NERSC supercomputers.

\begin{figure}[!htb] \includegraphics[width=0.457\textwidth]{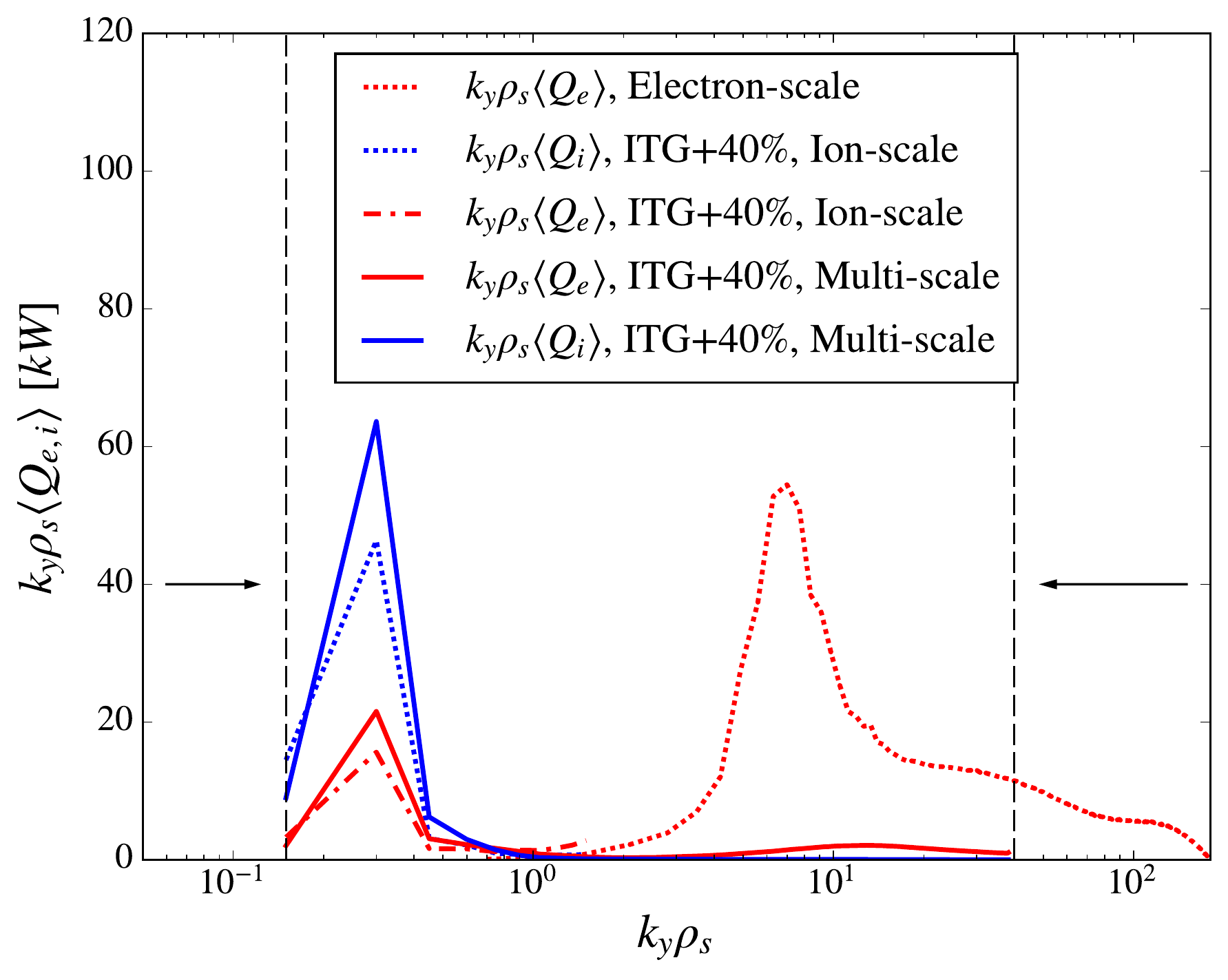}\caption{Nonlinear flux spectrum of multi-scale simulations (with simulation domain bounded by the dashed lines) and single-scale simulations. The area under the curves is proportional to the total heat flux carried at low- and high-k \cite{Goerler08PRL, Goerler08PoP}. Multi-scale simulations with ITG modes driven unstable by an increase in ITG by $40\%$ show that electron-scale heat flux via streamers is greatly reduced by large turbulent structures at the ion scales (see upper inset in Fig.~\ref{nonlinear}).} \label{fluxmulti}\end{figure}

Resolving the full ion and electron scales is computationally prohibitive (for the full extent of the electron and ion scales, see section \ref{gammarule}). We therefore conducted a series of single-scale simulations with a sequentially reduced box size. This was done to identify an affordable domain that still resolves the main physical behavior of the plasma. For example, at the ion scales, we found that we could increase $k_{y, \rm{min}}\rho_s=0.05 \rightarrow 0.15$ while maintaining the nonlinear heat flux to an accuracy of $\sim 10 \%$. Similarly, at the electron scales, we were able to reduce $k_{y, \rm{max}}\rho_s=180 \rightarrow 40$ with GyroLES techniques while maintaining a similar level of accuracy ($\sim15\%$) in the heat flux carried at the electron scales. This is significantly aided by the fact that most of the heat advection at the electron scales is carried by modes with $k_{y}\rho_s\approx8$. The flux-spectrum at the electron scales is plotted as the red dotted line in Figure \ref{fluxmulti}. In this type of plot, adapted from G\"orler and Jenko \cite{JenkoJap04, Goerler08PRL, Goerler08PoP, Waltz07}, the area under the curve roughly corresponds to the total heat flux carried by a range of wavenumbers. It is evident that, while the maximum growth rate in the linear flux spectrum is located at $k_y \rho_s\approx50$ (see Fig.~\ref{lineare}), the nonlinear heat flux is carried predominantly by streamers associated with ETG modes with wavenumbers in the vicinity of $k_y \rho_s\approx 8$ (see Fig.~\ref{fluxmulti}). 
This facilitates the above reduction in the resolved electron scales in the preparation for multi-scale simulations. At the ion scales, the heat flux is carried predominantly by modes with $k_{y}\rho_s>0.15$. We thus resolve both the electron and ion scales in a carefully selected domain of poloidal wavenumbers of $0.15 \leq k_y \rho_s \leq 40$ (see Fig.~\ref{fluxmulti}). Note that these nonlinear scans in simulation domain, while themselves computationally intensive, reduced the resource intensity of multi-scale simulations by a factor of~$\gtrsim10$, bringing them into the realm of the possible.

The nonlinear multi-scale simulations are performed with resolution $(n_x, n_y, n_z, n_{v_{\parallel}}, n_{\mu})=(512, 512, 16, 32, 18)$ and perpendicular box size $(L_x, L_y)=(75 \rho_s, 42 \rho_s)$. These simulations give the following qualitative results. First, we find that ETG-scale streamers co-exist with a zonal flow at ion scales when ITG modes are stable at nominal $\omega_{T_i}$. Second, we find that ETG-scale streamers are strongly sheared apart by ITG modes in the ITG+40\% scenario (see Fig.~\ref{fluxmulti}). In this scenario, a brief coexistence of ETG-streamers and an ion-scale zonal flow can also be observed intermittently before both are disrupted by ITG-streamers (see Supplementary Material).
Therefore, electron-scale transport does not contribute significantly to the total transport when ITG modes are highly unstable, such as when ITG+40\%. Thus, the heat-flux-matching {single}-scale simulation in Fig.~\ref{nonlinear} is representative of the {multi}-scale heat flux.

In summary, simulations at $\rho=0.80$ reproduce both the experimental ion and electron heat fluxes within the uncertainty of the CER data at the $1.6 \sigma$ level. Multi-scale simulations suggest that turbulent structures on the ion scales strongly disrupt the streamers found on the electron scales.

\section{Simulation Results at second radial position ($\mathbf{\rho=0.90}$) \label{edge}}

We now direct our attention further outwards to the near-edge region at $\rho=0.90$. We first carry out linear simulations to identify the linear mode spectrum, which is dominated by TEM/ETG turbulence. Subsequent nonlinear simulations show high sensitivity of the total heat flux to changes in the electron temperature gradient. This could be due to a hybrid ITG/TEM scenario that was not predicted by linear simulations. With the inclusion of $\bm{E} \times \bm{B}$ shear and an increase in the electron temperature gradient by 23\%, which is consistent with the experimental temperature data at the $1.3\sigma$ level, we are able to match the heat flux of the experiment. These results validate our gyrokinetic method and help push the gyrokinetic validation frontier closer to the edge region.

\subsection{Linear Simulations}
\begin{figure}[!htb] \includegraphics[width=0.457\textwidth]{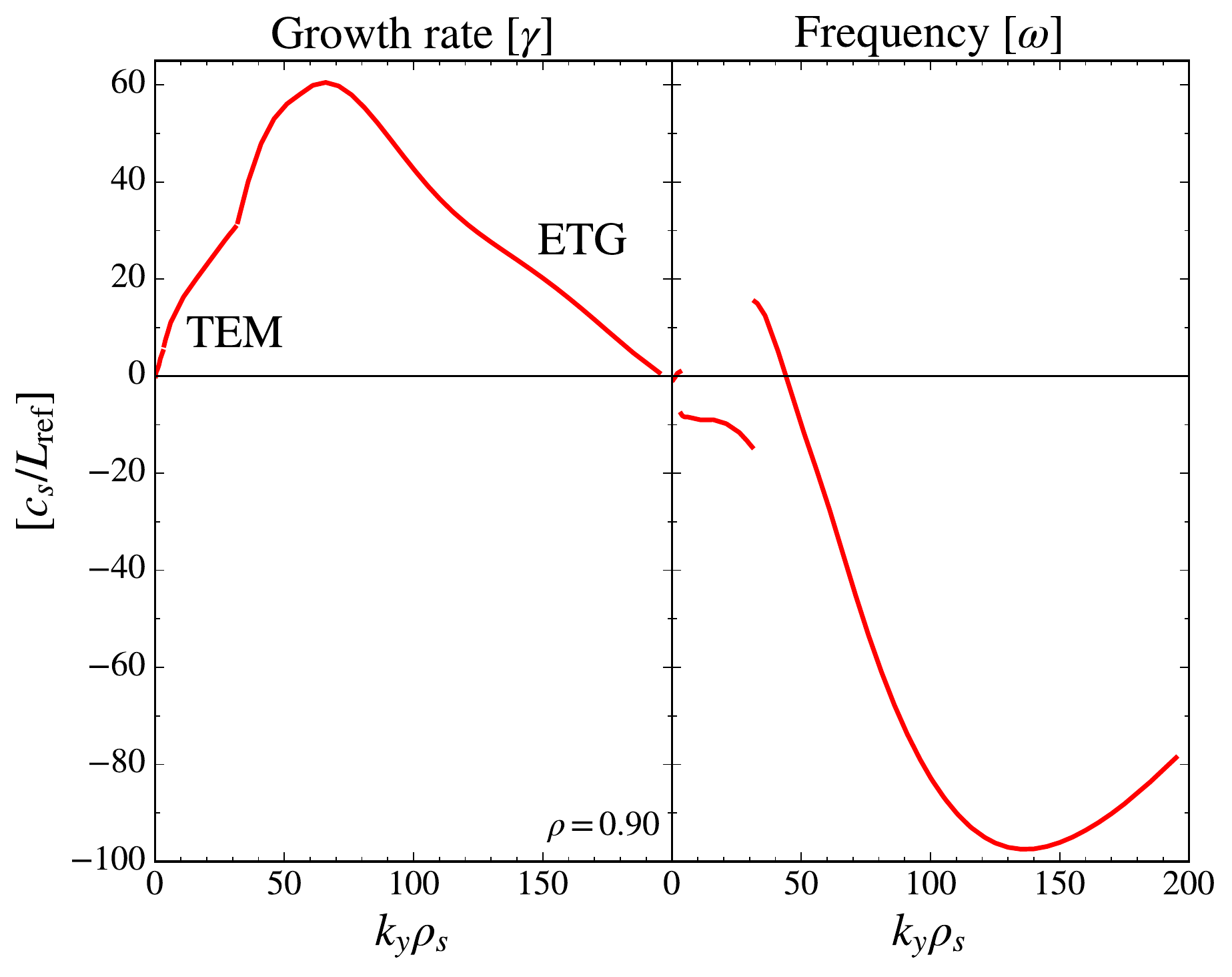}\caption{Linear growth rates and frequencies at $\rho=0.90$ on linear axes for nominal input parameters. We expect trapped electron modes (TEMs) at low-k and ETG modes at high-k.} \label{Lin_x90}\end{figure}

\begin{figure}[!htb] \includegraphics[width=0.457\textwidth]{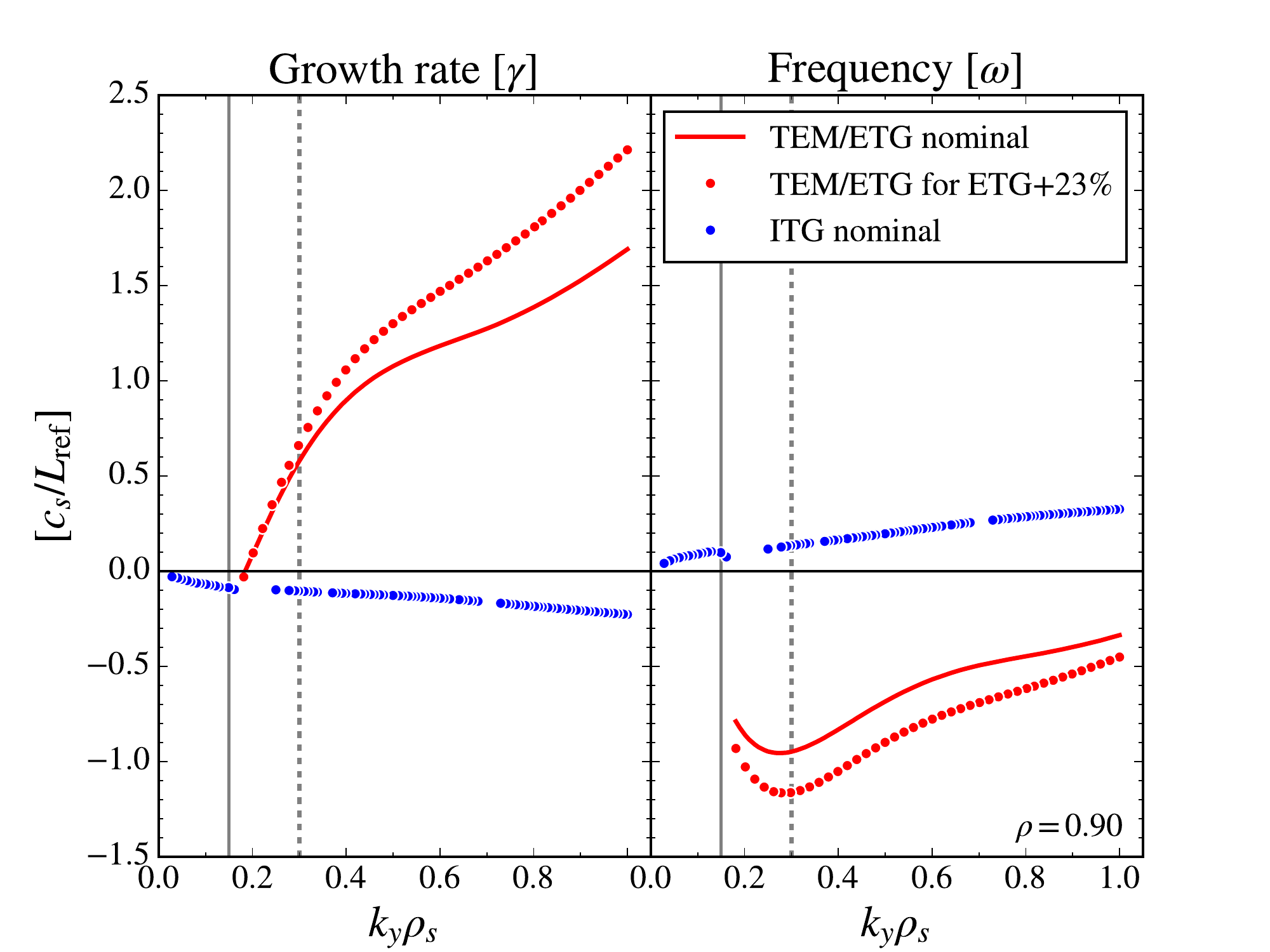}\caption{Linear growth rates and frequencies of TEM/ETG modes for nominal parameters (\emph{red line}) and with ETG increased by 23\% (\emph{red circles}). We also find subdominant and stable ITG modes (\emph{blue circles}). For comparison with nonlinear results below, the vertical lines highlight the position of the $k_y \rho_s = 0.15$ mode~(\emph{solid}) and the $k_y \rho_s = 0.30$ mode~(\emph{dashed}).}\label{sens09}\end{figure}

Linear simulations indicate that the main turbulent drive is carried by TEM/ETG modes at $\rho=0.90$ (see Fig.'s~\ref{Lin_x90} and \ref{sens09}). Note that this can be explained by the increased density gradient driving TEM turbulence closer to the plasma edge (see Fig.~\ref{neplot}). 
Figure \ref{sens09} shows the turbulent modes in low-wavenumber domain, which is most interesting for nonlinear simulations because the turbulent advection is most efficient at these large scales. To identify the subdominant modes that can play a role in nonlinear simulations, we employ the eigenvalue solver in GENE. Curiously, the subdominant mode is an ITG-type mode that is stable over all considered wavenumbers, which can be seen by its negative growth rates (see Fig.~\ref{sens09}). Moreover, we find that an increase in electron temperature gradient by $23\%$ further destabilizes the TEM/ETG modes. These linear simulations indicate that the TEM/ETG modes are likely to dominate in nonlinear simulations, with negligible effect of ITG modes.

\begin{figure}[!htb] \includegraphics[width=0.457\textwidth]{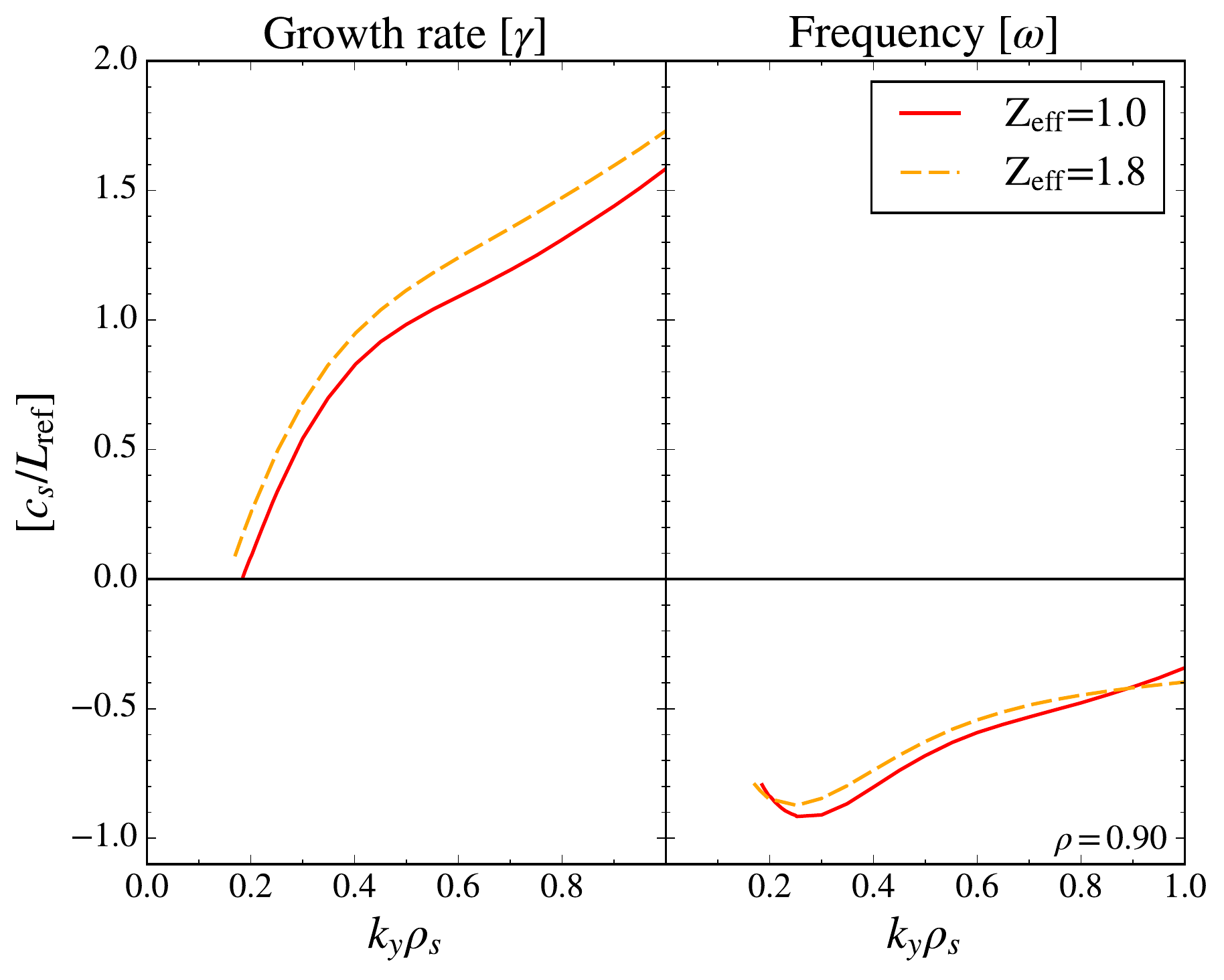}\caption{Linear growth rates and frequencies for a pure deuterium plasma (\emph{red}) and for a plasma with $Z_{\rm eff}=1.8$ due to a carbon impurity species (\emph{orange}). The effect of carbon impurities on growth rates is small and will be neglected in nonlinear simulations. } \label{Carbon09}\end{figure}

At $\rho=0.90$, the carbon impurity has a small effect on the linear growth rates compared to the pure deuterium plasma (see Figure \ref{Carbon09}). 
Therefore, to reduce computational complexity by approximately $30\%$, we carry out nonlinear simulations at $\rho=0.90$ with two rather than three particle species.

\subsection{Nonlinear Ion-scale Simulations}

Our linear simulations have shown that the growth rates at $\rho=0.90$ are sensitive to changes in $\omega_{T_e}$ (see Fig.~\ref{sens09}). This is due to the predominance of TEM/ETG-type modes with ITG modes subdominant and stable at nominal gradients. We thus study the sensitivity of nonlinear simulations to changes in $\omega_{T_e}$. 

The nonlinear ion-scale simulations are performed with resolution $(n_x, n_y, n_z, n_{v_{\parallel}}, n_{\mu})=(512, 64, 32, 32, 18)$ and perpendicular box size $(L_x, L_y)=(188\rho_s, 126\rho_s)$. Note that convergence tests found that a higher resolution was required for the ion scales at $\rho=0.90$ than at $\rho=0.80$. Physically, this is due to the need to resolve higher magnetic shear here (see Table \ref{params}). Moreover, the radial box size was increased because the simulated plasma was more susceptible to simulation boundary effects at $\rho=0.90$ than at $\rho=0.80$.  

We define the ion-scale domain as $0.05 \leq k_y \rho_s \leq 1.60$ and employ GyroLES techniques to avoid the unphysical build-up of free energy at $k_y \rho_s\simeq1.60$. The individual ion and electron heat channels are difficult to distinguish experimentally with current techniques due to the high collisionality at $\rho=0.90$, so that the observable here is the \emph{total} heat flux.
\begin{figure}[!htb] \includegraphics[width=0.457\textwidth]{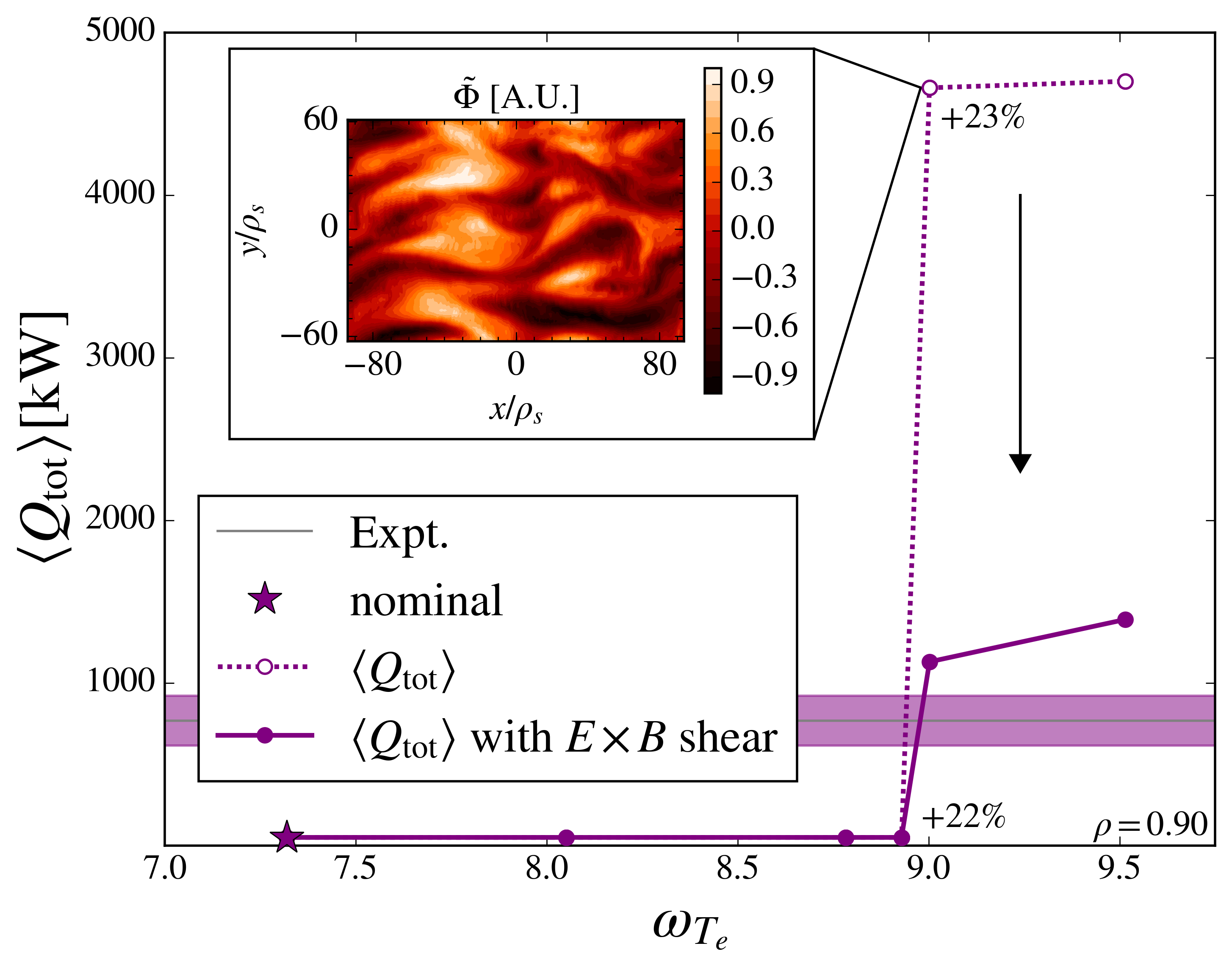}
\caption{Total heat flux as a function of the electron temperature gradient (ETG) in nonlinear simulations at $\rho=0.90$. Small changes in ETG from $+22\%$ to $+23\%$ show large changes in the heat flux, with high levels of turbulence for the $+23\%$ case (\emph{see inset}). 
The sensitivity to the electron temperature gradient could be caused by a hybrid TEM/ITG scenario investigated in the following figures. The sensitivity is tempered by the experimental $\bm{E}\times \bm{B}$ shearing rate.
} \label{NL09}\end{figure}

Nonlinear simulations with an increase in $\omega_{T_e}$ by up to $\sim30\%$ were performed.  Without the inclusion of experimental $\bm{E}\times \bm{B}$ shear, we found a saturated time-average flux of up to $\langle Q_{\rm tot}\rangle= 5$ MW (see Fig.~\ref{NL09}). Interestingly, we see high sensitivity to increases in $\omega_{T_e}$ between the $+22\%$ and $+23\%$ mark. The inset shows a contour plot of electrostatic potential fluctuations with high levels of turbulence at large scales for the $+23\%$ scenario without $\bm{E}\times \bm{B}$ shear. The observations persisted with an increase in the radial box size and an independent numerical test of the validity of GyroLES techniques. This indicates that there may be a physical origin for this high sensitivity of the heat flux to changes in ETG. 

\begin{figure}[!htb] \centering\includegraphics[width=0.457\textwidth]{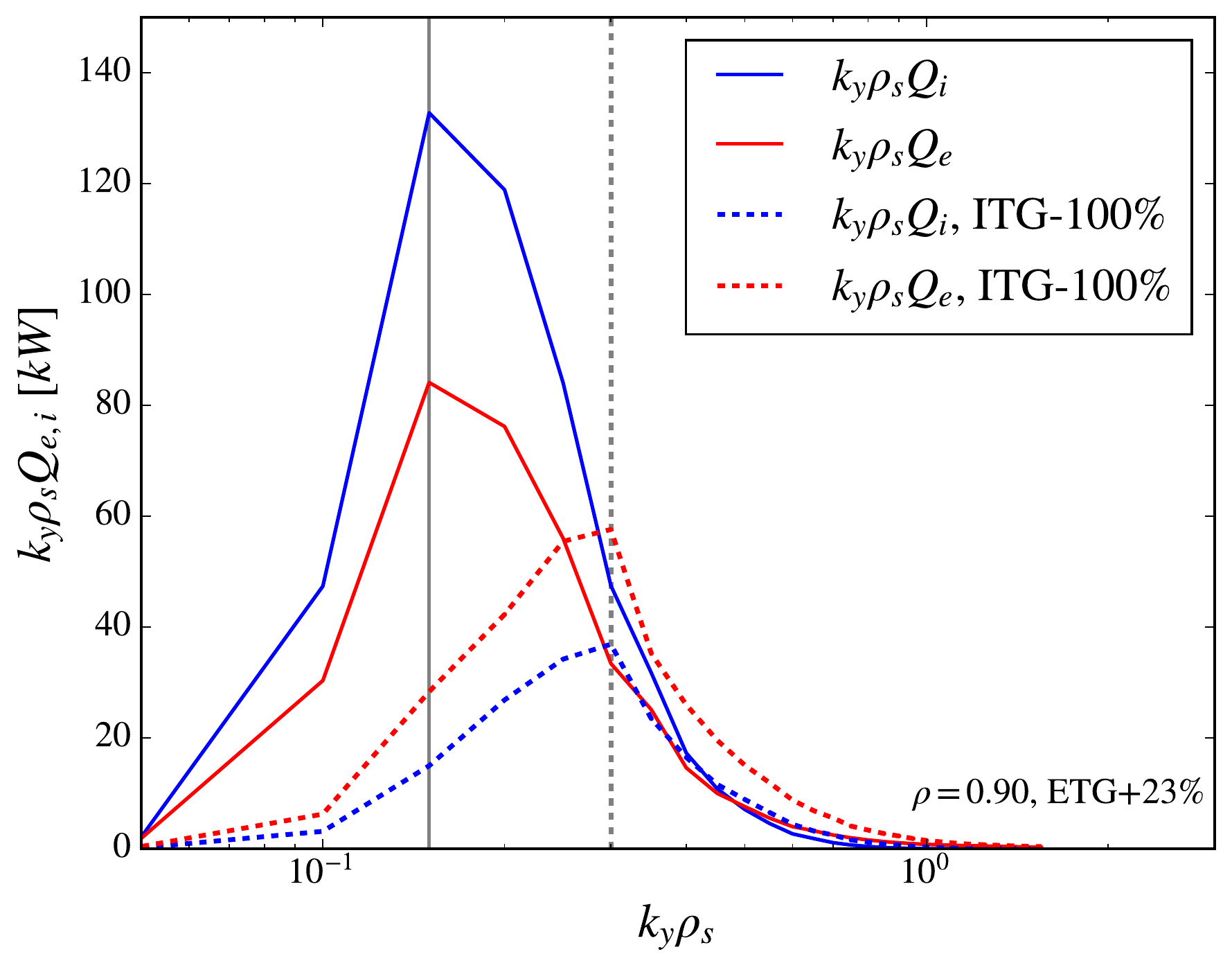}\caption{Heat-flux spectra of simulations with ETG increased by 23\% (ETG+23\%) and varying ITG. For nominal ITG, the majority of the heat flux is carried by the ions in the vicinity of $k_y \rho_s = 0.15$ (\emph{solid line}), indicating high ITG mode activity. For ITG reduced by 100\%, the majority of the heat flux is carried by the electrons in the vicinity of $k_y \rho_s = 0.30$ (\emph{dashed line}), indicating high TEM activity as originally predicted by linear simulations.}\label{NLFluxgradTe90}\end{figure}

The following analysis shows that TEM turbulence may nonlinearly destabilize the linearly stable ITG modes, leading to a hybrid ITG/TEM scenario in our nonlinear simulations. Recall that the ITG modes are \emph{stable} for all wavenumbers (see Fig.~\ref{sens09}). Moreover, the ETG/TEM branch with ETG increased by 23\% (ETG+23\%) is linearly unstable for $k_y\rho_s>0.18$. Based on these observations, we expected nonlinear simulations to show that the majority of the heat flux is carried by electrons in the vicinity of $k_y\rho_s \sim 0.30$. However, the heat-flux spectrum of nonlinear simulations with ETG+23\% shows that the majority of the heat flux is carried by ions in the vicinity of $k_y\rho_s \sim 0.15$ (see Fig.~\ref{NLFluxgradTe90}). This indicates high ITG mode activity, which was not predicted by linear simulations. To test the suspected ITG-dependence of the heat flux, we reduced the ion temperature gradient by 100\% and uncovered the purely TEM/ETG turbulence with electron-dominated heat flux carried predominantly at $k_y\rho_s \sim 0.30$ as originally expected. 

\begin{figure}[!htb] \centering\includegraphics[width=0.457\textwidth]{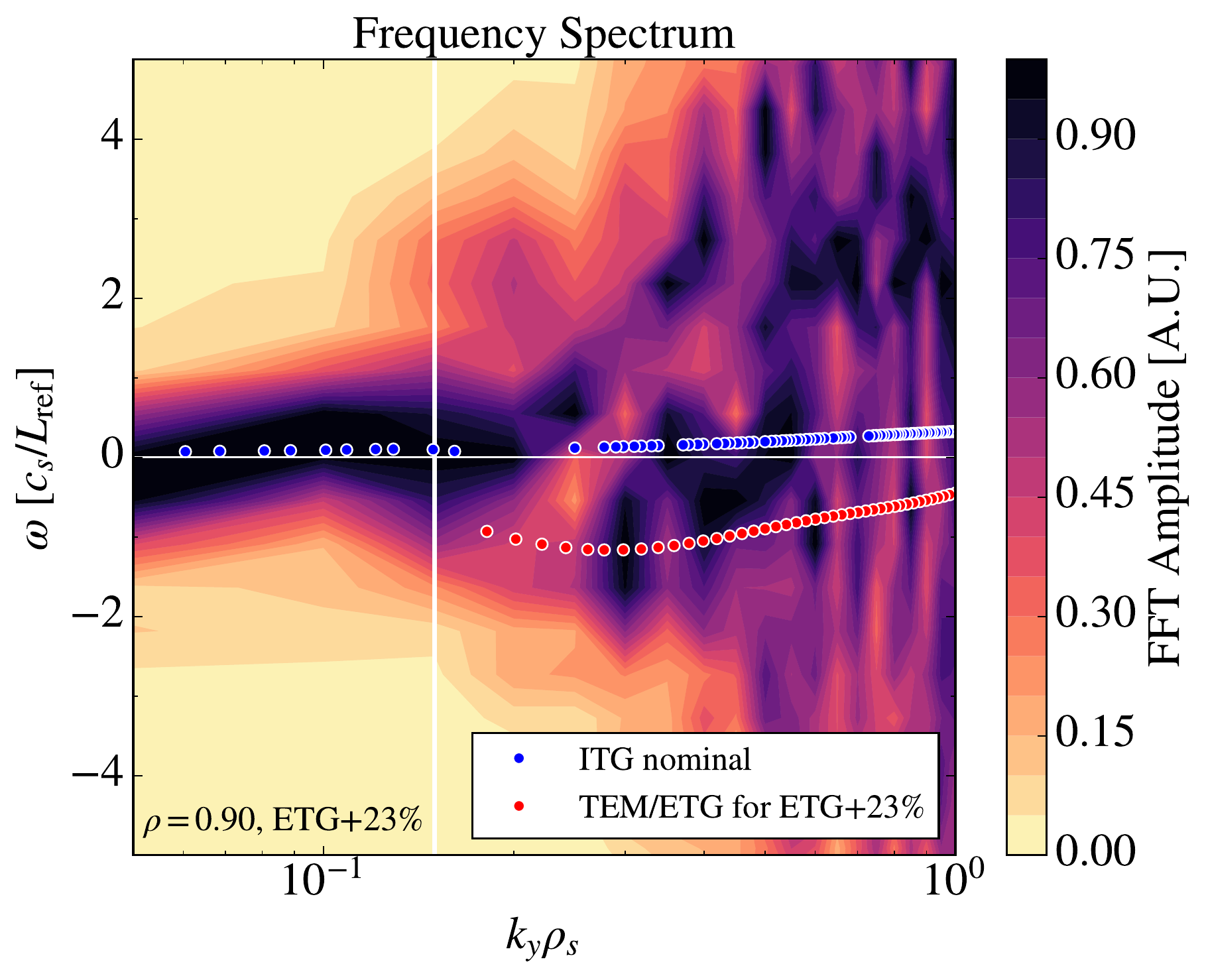}\caption{Frequencies of electrostatic potential fluctuations for the scenario with nominal ITG and ETG+23\%. The frequency of the nonlinear mode carrying most of the heat flux (\emph{vertical line}) is positive, which indicates that this is an ITG mode. However, in linear simulations all ITG modes (blue circles) are linearly stable. This suggests that the TEM/ETG instability (red circles), which has its nonlinear onset at ETG+23\%, may destabilize the ITG modes.}\label{NLFreqcontgradTe90}\end{figure}

To further identify the nature of nonlinear turbulence at $\rho=0.90$, we have plotted the frequency spectrum of electrostatic potential fluctuations (see Fig.~\ref{NLFreqcontgradTe90}). 
We find that the modes that carry most of the heat flux (i.e. $k_y \rho_s\sim0.15$) have positive frequency and therefore are associated with ITG turbulence. Moreover, the nonlinear frequencies at $k_y \rho_s= 0.15$ and $k_y \rho_s= 0.30$ are in good agreement with linear frequencies. This indicates that both TEM and ITG turbulence are active in nonlinear simulations. This seems to suggest that there exists a hybrid ITG/TEM scenario with nominal ITG and ETG+23\% (even though linear simulations did not predict this).

\begin{figure}[!htb] \centering\includegraphics[width=0.457\textwidth]{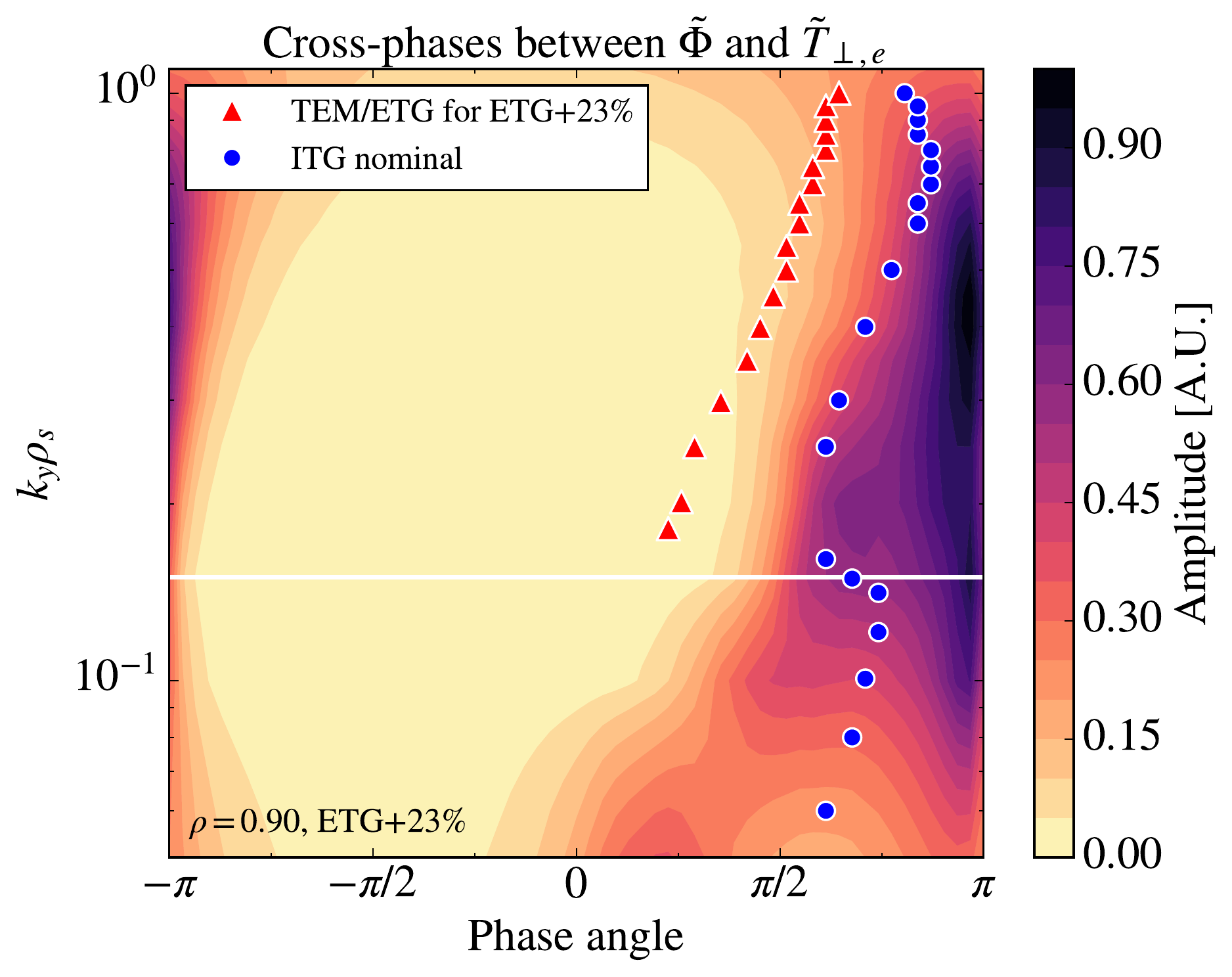}\caption{Cross-phases between fluctuations in the electrostatic potential and perpendicular electron temperature for the scenario with nominal ITG and ETG+23\%. At wavenumbers responsible for heat transport, $k_y\rho_s\sim0.15$ (\emph{horizontal line}), the nonlinear cross-phases agree with those of the linear ITG modes (\emph{blue circles}). For higher wavenumbers, the nonlinear cross-phases continue to agree with linear cross-phases associated with ITG modes rather than the TEM modes (\emph{red triangles}).}\label{Crossphase}\end{figure}

We now study the cross-phases between fluctuations in the electrostatic potential $\tilde{\Phi}$ and the perpendicular electron temperature $\tilde{T}_{\perp,e}$ (see Fig.~\ref{Crossphase}). 
For the wavenumbers carrying most of the heat flux, $k_y \rho_s \sim 0.15$, the nonlinear cross-phases agree with the linear cross-phases of the ITG modes. Moreover, at $k_y \rho_s \sim 0.15$ the nonlinear cross-phases are out of phase by approximately $\pi/2$, which is characteristic of significant electrostatic heat transport. For higher wavenumbers, the nonlinear cross-phases agree with the linear cross-phases of ITG modes rather than TEM/ETG modes. A similar hybrid ITG/TEM scenario was previously found in Ref. \citen{DTold13}, with the difference that both the dominant TEM and subdominant ITG modes were linearly unstable. Our results indicate a hybrid ITG/TEM scenario, with linearly subdominant and stable ITG modes carrying most of the heat flux. This scenario could also be relevant to spherical tokamaks, where ITG modes are more often linearly stable than in conventional tokamaks due to the lower aspect ratio~\cite{Morris99}.

One possible explanation for this scenario could be that the turbulent fluctuations of TEM turbulence nonlinearly excite the linearly stable ITG modes. This is reasonable by elimination, as there is insufficient linear drive and no alternative nonlinear drive for the ITG modes other than the TEM turbulence. Therefore, it appears that ETG+23\% is the critical gradient not only for TEM turbulence, but for hybrid ITG/TEM turbulence. This could explain the high heat-flux stiffness in our simulations at $\rho=0.90$ (see Fig.~\ref{NL09}). 

We now introduce electric field shear due to the L-mode $E_r$-well present in the edge region of most tokamaks. Generally, $E_r$-wells are much more pronounced in H-mode configurations, but are also present in L-mode plasmas \cite{Wagner} (see Fig.~\ref{WexB}). We find that a value of $\omega_{E\times B}=0.5$~$c_s/L_{\text{ref}}$, which is in the middle of the experimentally inferred range, is able to reduce the total heat flux approximately to the experimentally inferred values (see Fig.~\ref{NL09}). Therefore, the $\bm{E}\times \bm{B}$ shearing rate may be an important simulation parameter to accurately model the heat flux in the edge region of L-mode plasmas. We conclude that we are able to reproduce the total experimental heat flux at $\rho=0.90$ with an increase of $\omega_{T_e}$ by $\sim23\%$. This increase is within the combined uncertainty of the Thomson scattering data at the $1.3\sigma$ level (see Fig.~\ref{TiProf}).

Lastly, multi-scale effects may be present at $\rho=0.90$, but their investigation is computationally too expensive to fit within the scope of this work. Based on our experience with highly unstable ITG modes at $\rho=0.80$, we predict that the effect of electron-scale streamers would likely be strongly reduced by the highly unstable ITG/TEM turbulence found at the ion scales at $\rho=0.90$.

\section{Discussion}\label{disc}
To address an apparent shortfall problem\cite{Holland09, RhodesNF11}, recent validation exercises studying L-mode plasmas in Alcator C-Mod, ASDEX Upgrade, and DIII-D tokamaks have reduced fears that the shortfall is a universal feature of near-edge L-mode plasmas\cite{DTold13, Goerler14,HowardNF13, HowardNF16} (see section \ref{intro}). The results presented here are consistent with previous work that has been able to reproduce the experimental heat flux by changing input parameters close to their experimental uncertainty \cite{Goerler14, DTold13}. 

Results from multi-scale simulations with realistic Deuteron-electron mass ratio and geometry at $\rho=0.80$ have been presented. An early heuristic rule, found with pioneering multi-scale simulations\cite{JenkoJap04, Goerler08PRL, Goerler08PoP, Waltz07}, has suggested that ETG-modes can contribute experimentally relevant heat flux if $\gamma_{\rm ETG}^{\rm max}\left/ \gamma_{\rm ITG}^{\rm max}\right.\gg \sqrt{m_i/m_e}$. This rule of thumb was found using a reduced mass ratio and simplified geometry in the core, and is not expected to apply universally. Nevertheless, this rule appears to hold in recent multi-scale simulations using more realistic parameters with the GKV code \cite{GKV1, GKV2} and GYRO \cite{HowardPoP14, Maeyama}. In the present work, an example of the limit of applicability of this rule may have been found. Recall that linear simulations give $\gamma_{\rm ETG}^{\rm max}\left/ \gamma_{\rm ITG+40\%}^{\rm max}\right.=161$ (see section \ref{gammarule}). However, multi-scale simulations have qualitatively found very little ETG contribution to the overall heat flux with an increase in ITG by $40\%$ (see section \ref{nonlinsim}). Physically, large-scale turbulent structures of ITG-modes are able to shear ETG streamers apart. Thus, when ITG modes are highly unstable, they strongly reduce the flux carried by high-k modes. More recently, the condition $\text{Max}\left(\gamma_{\text{ETG}}/k_y\right)\geq\text{Max}\left(\gamma_{\text{ITG}}/k_y\right)\,$ has been identified as a simple test to predict possible multi-scale effects\cite{Staebler17}. This test has been validated with recent multi-scale simulations and seems to be less conservative than using only the ratio of maximum growth rates. For our linear simulations at $\rho=0.80$, this linear condition is only marginally satisfied for the ITG$+40\%$ scenario; namely, we get $\text{Max}\left(\gamma_{\text{ETG}}/k_y\right)/\text{Max}\left(\gamma_{\text{ITG}}/k_y\right)=1.16>1$ (see Fig.~\ref{Staebler}). Since this linear test is intended as a heuristic rule and not a hard-and-fast rule, our results of negligible multi-scale effects are not significantly at odds with this linear test. We therefore conclude that the recently proposed test appears to be more useful than relying only the ratio of maximum linear growth rates to instruct whether multi-scale simulations might be necessary. 

There are several limitations associated with our study. First, it became necessary to include the $\bm{E} \times \bm{B}$ shear for nonlinear simulations at $\rho=0.90$. As mentioned in section~\ref{meth}, the realistic shear in real geometry is given by the Hahm-Burrell formalism, which is not a flux function. For the purposes of our study, we assume that the flux-surface-average shear  given by the Waltz-Miller formalism is representative of the total shear effect. This is a common assumption in the simulation community~\cite{WaltzMiller99}. The more accurate Hahm-Burrell formalism has not been implemented in GENE yet. Nevertheless, our simplifying assumption does not affect the generality of our finding that $\bm{E} \times \bm{B}$ shear from the L-mode $E_r$-well is already important for simulations at $\rho=0.90$. 
Second, we have carried out multi-scale simulations in the near-edge for the first time. These simulations would not be possible without simplifying assumptions. For instance, we have relied on GyroLES techniques to replace unresolved dissipation with a model. Moreover, we have relied on a large number of single-scale simulations to empirically determine the validity of our multi-scale simulation parameters. However, multi-scale convergence tests may be helpful to test the sensitivity of multi-scale effects to simulation parameter changes. Since these are computationally expensive they are beyond the scope of the present work. Third, note that particularly the ion temperature data used for this work has a large uncertainty associated with it, likely due to low carbon density (see Fig.~\ref{TiProf}). We have estimated the relevant statistical uncertainties for the flux-matching temperature gradients to be $\sigma_{\text{ITG, stat}}\sim15\%$ and $\sigma_{\text{ETG, stat}}\sim8\%$ (see Appendix). We have estimated reasonable systematic uncertainties as $\sigma_{\text{sys}}\sim~10\%$. As a result of these error estimates, we find that our flux-matching gradients of ITG+40\% and ETG+23\% both fall within $\leq2\sigma_{\text{stat}}+1\sigma_{\text{sys}}$. It is worth repeating here that only slightly more than two thirds ($68\%$) of normally distributed measurements fall within $1\sigma_{\text{stat}}$, while most ($95\%$) fall within $2\sigma_{\text{stat}}$ and nearly all ($99.7\%$) fall within $3\sigma_{\text{stat}}$. Therefore, our deviations from the experimental measurements of $\leq2\sigma_{\text{stat}}+1\sigma_{\text{sys}}$ are acceptable from a purely statistical perspective and sufficient for gyrokinetic validation. Nonetheless, there may be limitations in our validation method that exclude some relevant physics. For example, we are only considering a local flux-tube domain for our simulations and we are assuming a purely Maxwellian background distribution. While we have ruled out multi-scale effects, using a global model or including non-Maxwellian fast-ions could contribute sufficient missing physics to affect our conclusions. Including these phenomena is beyond the scope of the present work.

\section{Summary}\label{sum}
We have presented results from a study of a DIII-D L-mode plasma in the near-edge. At $\rho=0.80$, the radial ion flux is quenched by strong poloidal zonal flows for nominal input parameters. In the ITG+40\% scenario, nonlinear single-scale simulations give remarkable agreement with both the ion and electron heat fluxes of the experiment. This change in gradient is compatible with the combined statistical and systematic uncertainty in the ion temperature gradient at the $1.6 \sigma$ level (see Fig.~\ref{TiProf}). At the electron-scales, radially elongated streamers are found to carry significant electron heat flux that is comparable to the experiment. This motivates multi-scale simulations, which were carried out for the first time in the near-edge with realistic mass ratio and geometry. Results suggest that the highly unstable ITG modes in the flux-matched ion-scale simulations strongly suppress turbulent transport at the electron-scales. Therefore, single-scale simulations are sufficient to match the experimentally inferred heat flux by changing the ion temperature gradient within the uncertainty of the experiment at $\rho=0.80$. At $\rho=0.90$, nonlinear simulations uncover a hybrid ITG/TEM scenario, which was not predicted by linear simulations. Moreover, nonlinear simulations are able to match the total experimental heat flux in the ETG+23\% scenario when $\bm{E} \times \bm{B}$ flow shear (as evaluated from Doppler Backscattering measurements) is taken into account. This is consistent with the combined uncertainty of the electron temperature measurements at the $1.3\sigma$ level. Therefore, our primary conclusion is that gyrokinetic simulations are able to match the heat-flux in the near-edge of the L-mode plasma by varying input parameters close to their experimental uncertainties at $\rho=0.8$ and $\rho=0.9$. 

\section{Future Work}\label{future}
The present work invites several avenues for future research. 
For instance, future work could study the nature of multi-scale effects for the ITG$+3.5\%$ scenario: When ITG modes are marginally unstable, previous multi-scale simulations in the core have found that (i) ETG streamers can contribute experimentally significant transport at small scales \cite{HowardNF13, HowardNF16} and (ii) ETG streamers can dampen poloidal zonal flows and enhance ion-scale transport \cite{Maeyama}. Similar effects might be found with our simulations in the near-edge when ITG modes are brought close to marginal stability. Building upon the results presented here, future work could quantify the effect of multi-scale interactions in the near-edge of L-mode plasmas.

The main ion temperature profile is often assumed to be equal to the impurity ion temperature measured with the CER diagnostic (see section \ref{exp}). This could introduce systematic errors. However, recent diagnostic development at DIII-D is now able to extract the main ion temperature directly and is currently quantifying this known source of uncertainty \cite{Haskey16}. A main ion CER (MICER) diagnostic has been developed to study the deuterium ion ($D^+$) charge exchange signal.
In order to test the conclusions of this work, future work could study a similar L-mode discharge using data from the more precise and more accurate MICER diagnostic currently under development at DIII-D \cite{Haskey16}. 

Closer to the edge region, global simulations are likely required, as the shearing rate changes substantially within a narrow radial region just inside the last closed flux surface. Resistive ballooning modes can also be expected to potentially contribute to thermal edge transport\cite{Bourdelle12, Cohen13, Rafiq14}. Future work could extend the present gyrokinetic simulations closer to the edge of plasmas just before the L-H transition \cite{Lothar, Schmitz17}.

\section*{Supplementary Material}
See the \href{https://doi.org/10.1063/1.5052047}{supplementary material} for a movie of the electrostatic potential fluctuations $\tilde{\Phi}$ [A.U.] in multi-scale simulations with resolution $(n_x, n_y, n_z, n_{v_{\parallel}}, n_{\mu})=(512, 512, 16, 32, 18)$ and box size $(L_x, L_y)=(75 \rho_s, 42 \rho_s)$. This movie qualitatively shows the formation of large turbulent structures associated with the ion scales and the disruption of the small-scale horizontal streamers associated with the electron scales. 

\section*{Acknowledgements}
The authors wish to thank Tobias G\"orler, Nathan Howard, Chris Holland, Walter Guttenfelder, Craig Petty, Punit Gohil, Wayne Solomon, Martin Weidl and Qingjiang Pan for helpful questions and comments. This work is supported by the US Department of Energy (DOE) under award number DE-SC0016073. The computational effort was conducted at NERSC, a DOE Office of Science User Facility supported under contract DE-AC02-05CH11231. The experimental work at DIII-D was supported by the US DOE under contract numbers DE-FG02-08ER54984 and DE-FC02-04ER54698.

\section*{Disclaimer}
This report was prepared as an account of work sponsored by an agency of the United States Government. Neither the United States Government nor any agency thereof, nor any of their employees, makes any warranty, express or implied, or assumes any legal liability or responsibility for the accuracy, completeness, or usefulness of any information, apparatus, product, or process disclosed, or represents that its use would not infringe privately owned rights. Reference herein to any specific commercial product, process, or service by trade name, trademark, manufacturer, or otherwise, does not necessarily constitute or imply its endorsement, recommendation, or favoring by the United States Government or any agency thereof. The views and opinions of authors expressed herein do not necessarily state or reflect those of the United States Government or any agency thereof.

\section*{Appendix: Error analysis\label{app}}
\setcounter{equation}{0}
\renewcommand{\theequation}{A\arabic{equation}}
In order to quantify the statistical uncertainty of the ion temperature data, a $\chi^2$ fitting procedure is employed. As theoretical model for the temperature, we use a polynomial of the lowest order ($n=3$) that can pass through all data points within their uncertainty, namely 
\begin{equation}\label{A1}
T_{i, \rm{th}}(\rho)=a(\rho-0.80)^3+b(\rho-0.80)^2+c(\rho-0.80)+T_{i,0}\,,
\end{equation}
where $T_{i,0}$ is the temperature at $\rho=0.80$. Note that this model does not rely on the underlying physics that generates this profile, but rather represents a suitable functional form. We assume that experimental temperature measurements $T_{i, \rm{exp}}(\rho)$ are independent and normally distributed, inviting a $\chi^2$ statistic 
\begin{equation}\label{A2}
\chi^2(\theta)=\sum_i \frac{\left[T_{i, \rm{th}}(\rho_i)-T_{i, \rm{exp}}(\rho_i)\right]^2}{\sigma_i^2}\,,
\end{equation}
where $\theta=(a,b,c,T_{i,0})$ are the free parameters. We want to quantify the error associated with the temperature gradient, 
\begin{equation}\label{A3}
\nabla T_{i, \rm{th}}(\rho=0.80)=c\,.
\end{equation}
We use Bayes' theorem to find the probability density function (PDF) of the free parameter $c$,\begin{equation}\label{A4}
p(c|\theta)=\frac{\int_{-\infty}^{\infty} \,da \int_{-\infty}^{\infty}\,db\int_{0}^{\infty}{\rm exp}(-{\chi^2}/{2})\,dT_{i,0} }{\int_{-\infty}^{\infty} \,da \int_{-\infty}^{\infty}\,db\int_{-\infty}^{\infty}\,dc\int_{0}^{\infty}{\rm exp}(-{\chi^2}/{2})\,dT_{i,0} }\,.
\end{equation}
This gives a best-fit value of $c \approx 1.0\pm0.15\,(\rm{stat.})$ keV or $\sigma_{\text{ITG, stat}}\sim~15\%$. 
One known weakness of the $\chi^2$ fitting procedure is that it assumes the theoretical model is correct. To account for possible systematic errors in the model assumptions, e.g. a polynomial fit to the data, and the experimental setup, e.g. neutral beam halo effects \cite{Delabie11} or Zeeman splitting of the (carbon) impurity line used for CER \cite{Haskey18}, we allow for a systematic uncertainty of $\sigma_{\text{ITG, sys}}\sim 10\%$. Conservatively combining these two estimated uncertainties linearly gives an uncertainty of
\begin{equation}
\sigma_{\text{ITG}} \sim 25\%\,.
\end{equation}
Thus, the logarithmic ion temperature gradient at $\rho=0.80$ to one standard deviation is approximately $\omega_{T_i} = 2.78\pm0.70$.

The above method is also used to estimate the uncertainty in the electron temperature gradient at $\rho=0.90$. We find that a third order polynomial fits the electron temperature data well in the range $0.33<\rho<1.00$, which supplies a sufficiently large data set for our error analysis. Applying equations (\ref{A1})-(\ref{A4}) to this scenario gives a statistical uncertainty of $\sigma_{\text{ETG, stat}} \sim 8\%$ for the electron temperature gradient at $\rho=0.90$. However, the temperature profile of a tokamak plasma is subject to microscopic fluctuations due to predator-prey dynamics between gradient-driven microturbulence and zonal flows\cite{Lothar, Schmitz17}. While these dynamics have been observed at large scales, current experimental techniques do not yet capture these dynamics on small scales. Therefore, a systematic uncertainty in the temperature gradients is physically motivated. As before, allowing for a systematic uncertainty of $\sigma_{\text{ETG, sys}}\sim 10\%$ gives a linearly combined uncertainty of 
\begin{equation}
\sigma_{\text{ETG}} \sim 18\%\,.
\end{equation}
Therefore, an increase in the electron temperature gradient by $23\%$ is within $\sim 1.3 \sigma$ of the experimental value.

%

\end{document}